\documentclass[fleqn,usenatbib,onecolumn,useAMS]{mnras}

\usepackage{newtxtext,newtxmath}

\usepackage[T1]{fontenc}

\DeclareRobustCommand{\VAN}[3]{#2}
\let\VANthebibliography\thebibliography
\def\thebibliography{\DeclareRobustCommand{\VAN}[3]{##3}\VANthebibliography}

\usepackage{graphicx}	
\usepackage{amsmath}	

\newcommand{\de}{\mathrm{d}}

\newcommand{\derivp}[2]{\dfrac{\partial #1}{\partial #2}}
\newcommand{\derivpp}[2]{\dfrac{\partial^{2} #1}{\partial #2^{2}}}

\newcommand{\bvec}[1]{\boldsymbol{#1}}
\newcommand{\dint}{\displaystyle\int}
\newcommand{\dsum}{\displaystyle\sum}

\newcommand{\eqsali}[1]{\begin{equation}\begin{aligned}#1\end{aligned}\end{equation}}

\pdfoutput=1

\title[alicptXcsst]{Forecast of cross-correlation of CSST cosmic shear tomography with AliCPT-1 CMB lensing}

\author[Z. Wang et al]{
Zhengyi Wang$^{1,2}$,
Ji Yao$^{3}$\thanks{E-mail: ji.yao@shao.ac.cn},
Xiangkun Liu$^{4}$,
Dezi Liu$^{4}$,
Zuhui Fan$^{4}$,
Bin Hu$^{1,2}$\thanks{E-mail: bhu@bnu.edu.cn}
\\
Institute for Frontier in Astronomy and Astrophysics, Beijing Normal University, Beijing, 102206, China$^{1}$\\
Department of Astronomy, Beĳing Normal University, Beĳing 100875, China$^{2}$ \\
Shanghai Astronomical Observatory, Nandan Road 80, Shanghai 200030, China$^{3}$\\
South-Western Institute for Astronomy Research, Yunnan University, Kunming 650500, People's Republic of China$^{4}$\\
}

\date{Accepted XXX. Received YYY; in original form ZZZ}

\pubyear{2015}

\begin{document}
\label{firstpage}
\pagerange{\pageref{firstpage}--\pageref{lastpage}}
\maketitle

\begin{abstract}
We present a forecast study on the cross-correlation between cosmic shear tomography from the Chinese Survey Space Telescope (CSST) and CMB lensing from Ali CMB Polarization Telescope (AliCPT-1) in Tibet. The correlated galaxy and CMB lensing signals were generated from Gaussian realizations based on inputted auto- and cross-spectra. To account for the error budget, we considered the CMB lensing reconstruction noise based on the AliCPT-1 lensing reconstruction pipeline; shape noise of the galaxy lensing measurement; CSST photo-$z$ error; photo-$z$ bias; intrinsic alignment effect, and multiplicative bias. The AliCPT-1 CMB lensing mock data were generated according to two experimental stages, namely the ``4 modules*yr'' and ``48 modules*yr'' cases. We estimate the cross-spectra in 4 tomographic bins according to the CSST photo-$z$ distribution in the range of $z\in[0,4)$. After reconstructing the pseudo-cross-spectra from the realizations, we calculate the signal-to-noise ratio (SNR). By combining the 4 photo-$z$ bins, the total cross-correlation SNR$\approx15$ (AliCPT-1 ``4 modules*yr'') and SNR$\approx22$ (AliCPT-1 ``48 modules*yr''). Finally, we study the cosmological application of this cross-correlation signal. Excluding intrinsic alignment (IA) in the template fitting would lead to roughly a $0.6\sigma$ increment in $\sigma_8$ due to the negative IA contribution to the galaxy lensing data. For AliCPT-1 first and second stages, the cross-correlation of CSST cosmic shear with CMB lensing gives errors on the clustering amplitude $\sigma_{\sigma_8}=^{+0.043}_{-0.038}$ or $\sigma_{S_8}=\pm 0.031$ and $\sigma_{\sigma_8}=^{+0.030}_{-0.027}$ or $\sigma_{S_8}=\pm 0.018$, respectively.
\end{abstract}
\section{Introduction}

Weak gravitational lensing (WL) is a powerful tool in constraining cosmology  \citep{Refregier:2003ct,Mandelbaum:2017jpr,Blake:2020mzy,Jullo:2019lgq,Zhang:2007nk}, however, the so-called ``$S_8$ tension'' between WL and cosmic microwave background (CMB) anisotropy observations prevent us from using their synergy \citep{Planck:2018lbu,Planck:2018vyg,DES:2021wwk,Heymans:2020gsg,KiDS:2020suj,HSC:2018mrq,Hamana:2019etx}. Whether this tension is due to new physics beyond the standard $\Lambda$CDM model will require us to carefully address all kinds of possible systematics \citep{Wright:2020ppw,Yao:2020jpj,Yao:2017dnt,Kannawadi:2018moi,mead2021hmcode,DES:2021vln}. The cross-correlations between different tracers are important tools in investigating such problems, as they are immune to many systematics and can bring extra cosmological information.

Cross-correlations of galaxy surveys with CMB lensing provide a powerful way to probe the large-scale structure (LSS) of the Universe (\citealt{robertson2021strong}; \citealt{zhang2021transitioning}; \citealt{omori2022joint}). CMB lensing can be reconstructed by observing temperature and polarization anisotropies, it allows us to reconstruct a map of the integrated over-the-line-of-sight overdensity of intervening matter \citep{Hu:2001kj,Okamoto:2003zw}. Galaxy image surveys provide a trace of LSS, WL can be detected by capturing the images of a large sample of galaxies, usually called source galaxies, and performing shape measurements that can then be analyzed statistically \citep{camacho2021cosmic,Mandelbaum:2017jpr,Bartelmann:1999yn}. 
The correlations between the CMB lensing and cosmic shear\footnote{WL generated by LSS.} have been extensively measured, however with low significance, due to the limit of the current observations \citep{Robertson:2020xom,DES:2015wix,Harnois-Deraps:2016huu,DES:2018nnc,Hand:2013xua,Marques:2020dsb,Liu:2015xfa,Harnois-Deraps:2017kfd,POLARBEAR:2019phb,Singh:2016xey}.

Several CMB experiments have reconstructed the CMB lensing signals, such as Planck \citep{Planck:2018lbu}, ACT \citep{Mallaby-Kay:2021tuk,Darwish:2020fwf} and SPTpol \citep{DES:2017fyz,Wu:2019hek}. The next stage CMB observations like CMB-S4 \citep{abazajian2016cmb} and Simons Observatory \citep{2019JCAP...02..056A} will also do these observations with high significance. 
AliCPT-1 \citep{li2018tibet,Li:2017drr,Ghosh:2022mje,salatino,2020SPIE11453E..2AS} is the first Chinese CMB experiment aiming for high-precision measurement of Cosmic Microwave Background B-mode polarization. The telescope, currently under deployment in Tibet, will observe in two frequency bands centered at 90 and 150 GHz. As predicted in Refs. \citep{liu2022forecasts,disandai}, AliCPT-1 has the potential to reconstruct the CMB lensing with a relatively high signal-to-noise ratio.
Considering two different stages of AliCPT-1, namely ``4 modules*yr''\footnote{The unit denotes for the accumulated data length.} and ``48 modules*yr'', we are able to measure the lensing signal at $15\sigma$ and $31\sigma$ significance, respectively. For simplicity, in this paper we quantify the difference between the two stages in terms of the length of the data vector\footnote{he data vector refers to the time-ordered data (TOD) of CMB. The ``48 modules*yr'' is obtained from 4 seasons according to our current detector upgrade plan. The ``4 modules*yr'' is assuming 4 modules observe for 1 season.}. The latter is 12 times longer than the former. Hence, the latter can suppress the statistical noise by a factor of $\sqrt{12}$ \cite{liu2022forecasts}.

High-resolution imaging in optical and infrared bands is one of the major scientific tasks of many galaxy surveys, such as Rubin Observatory \citep{abell2009lsst}, Euclid \citep{laureijs2011euclid} and Roman Space Telescope \citep{spergel2015wide}.
The Chinese Survey Space Telescope (CSST) \citep{zhanhu,Gong:2019yxt,2018MNRAS.480.2178C} project also belongs to the stage-IV galaxy surveys. It is a 2-meter space telescope in the same orbit as the China Manned Space Station. he wide field survey will cover $17,500$ deg$^2$ sky area in about 10 years with field-of-view (FOV) 1.1 deg$^2$. The photometric filters are in $NUV, u, g, r, i, z$, and $y$ bands. The point source $5\sigma$ limiting magnitudes in $g$ and $r$ bands can reach 26 (AB mag) or higher. It will simultaneously perform both photometric imaging and slitless grating spectroscopic surveys with high spatial resolution $0.15^{''}$ ($80\%$ energy concentration region) and wide wavelength coverage. There are seven photometric imaging bands and three spectroscopic bands covering $255\sim1000$ nm. 
The mean number density of galaxies is about 20 galaxies per arcmin$^2$, which is triple the current stage-III galaxy survey. Hence, CSST is an excellent instrument for WL studies.

This paper presents a study on the cross-correlation between cosmic shear from CSST and CMB lensing from AliCPT-1. We built a simulation pipeline using Gaussian realization of signal and noise spectra for this purpose. For cosmic shear, we consider three types of noises or biases, namely photo-$z$ error, galaxy shape noise as well as intrinsic alignment. For CMB lensing, we consider the noise from the disconnected primary CMB. 
We explore the impact of systematics, with a main focus on the photo-$z$ and intrinsic alignment, and different types of observational noise.
Next, we estimate the standard Pseudo-$C_\ell$ spectrum estimation and compute the corresponding signal-to-noise ratio (SNR). Lastly, we investigate the cosmological implications of these cross-correlated signals.

This paper is organized as follows. In Section \ref{sec2} we will introduce the basics of CMB lensing and WL. In Section \ref{sec3}, we will present the method of simulating the CMB lensing $\kappa$ maps and the cosmic shear $\gamma$ maps, including both the cosmological signal and the systematic effects. In Section \ref{sec4} we will present the cross-correlation pseudo-$C_\ell$ measurements method and describe the covariance matrix, computing SNR. In Section \ref{sec5} we will present the cosmological parameter constraint results based on the simulated maps.

\section{Theoretical Modelling}\label{sec2}

\subsection{CMB lensing}

CMB lensing signal reconstructed from CMB temperature and polarization maps traces the integrated matter field along the line-of-sight direction from the current observer to the last scatter surface with redshift $z_*\simeq1100$
\eqsali{
\kappa_{\mathrm{CMB}}(\bvec{\theta})=\dint^{z_{*}}_{0}\de z q_\mathrm{CMB}(z)\delta(\chi(z)\bvec{\theta},z)\;,
}
where lensing kernel $q_\mathrm{CMB}(z)$ reads
\eqsali{
q_\mathrm{CMB}(z)=\dfrac{3\Omega_{m0}}{2c}\dfrac{H^{2}_{0}}{H(z)}(1+z)\chi(z)\;,\label{qcmb}
}
$\chi$ is the comoving distance, $H(z)$ is the Hubble parameter. 
As is well known, the maximum efficiency of lensing occurs when the lenses are positioned halfway between the source and the observer. In the case of CMB lensing, the background light originates from the last scattering surface. Consequently, the lensing efficiency gradually increases from low redshift and plateaus after $z\approx1$ and peaks at $z\simeq 2$. This plateau remains constant until high redshift values. Unlike cosmic shear, which necessitates tomographic analysis based on the distribution of source galaxies, CMB lensing signals can compress a vast majority of cosmological matter distribution information into a single sphere. 
For CMB lensing reconstruction, we adopt the Planck 2018 lensing paper \citep{Planck:2018lbu,Carron:2017mqf} formalism, which closely resembles the original Hu-Okamoto formalism \citep{Hu:2001kj,Okamoto:2003zw}. 
Since lensing creates correlations between different multipoles, the quadratic estimator utilizes these induced correlations in an almost optimal and unbiased way. The quadratic estimator spectrum captures the sought-after signal but also unavoidably includes Gaussian reconstruction noise sourced by both the CMB and instrumental noise (N0 bias) and non-primary couplings of the connected 4-point function\citep{Kesden:2003cc} (N1 bias). In the AliCPT-1 experimental setup\citep{liu2022forecasts}, the former is completely dominant compared to the latter two biases. Therefore, we only consider the N0 term in the n  oise budget of CMB lensing in this work.

\subsection{Cosmic shear}

The weak lensing effect of source galaxies dues to the intervening large-scale structure is called cosmic shear. The corresponding lensing potential $\phi(\boldsymbol{\theta})$ is defined as

\eqsali{
\phi(\boldsymbol{\theta})=-\dfrac{2}{c^2}\displaystyle\int^{\chi_{*}}_{0}\de \chi\dfrac{\chi_{*}-\chi}{\chi_{*}\chi}\Psi(\boldsymbol{\theta},\chi)\label{lensing_potential}
}
where $\Psi(\boldsymbol{\theta},\chi)$ is the 3D Weyl gravitational potential. It induces a distortion of the shape and provides a mapping from lens plane position $\bvec\theta$ to the source plane position $\bvec\beta$. The local properties of a gravitational lens are characterized by the Jacobian matrix $\mathcal{A}$ of the mapping given by \citep{BARTELMANN2001291}

\eqsali{        \mathcal{A}=\derivp{\bvec\beta}{\bvec\theta}
        =\delta_{ij}+\dfrac{\partial^2\phi}{\partial\theta_i\partial\theta_j}
        =
                \left[
    \begin{array}{cc}
        1-\kappa-\gamma_1 & -\gamma_2 \\
        -\gamma_2 & 1-\kappa+\gamma_1
    \end{array}
\right]\label{Jacobian}
}
with
\eqsali{
\kappa=-\dfrac{1}{2}\nabla^2\phi ,\quad \gamma_1=\dfrac{1}{2}\left(\derivpp{\phi}{\theta_1}-\derivpp{\phi}{\theta_2}\right) ,\quad \gamma_2=\dfrac{\partial^2\phi}{\partial\theta_1\partial\theta_2}\;,\label{kg1g2}
}
where $\kappa$ is the lensing convergence and $\gamma=\gamma_1+i\gamma_2$ is the complex lensing shear field.

Due to the intrinsic ellipticity of the source galaxies, the observed galaxy shape is actually the mixture of the cosmic shear and the intrinsic ellipticity \citep{bonnet1995statistical}. 
Assuming 
$\kappa \ll 1$ and $\gamma_1,\gamma_2 \ll 1$, up to the first order, the lensed ellipticity reads
\eqsali{\varepsilon = \varepsilon_s + \gamma\;,
}
where $\varepsilon_s$ is the intrinsic ellipticity of the source galaxy.
Using the Poisson equation one can write the convergence in terms of the density perturbation $\delta(\bvec\theta,z)$ 
\eqsali{
\kappa_i(\bvec\theta)=\dint^{z_*}_0\dfrac{c\de z}{H(z)}q_i(z)\delta(\bvec\theta,z)\;,
}
where lensing efficiency $q_i(z)$ in cosmic shear measurement is defined 
\eqsali{
q_i(z)=\dfrac{3\Omega_{m0}}{2}\dfrac{H^2_0}{c^2}(1+z)\chi(z)\dint^{z_*}_{z}n_i(z')\dfrac{\chi(z')-\chi(z)}{\chi(z')}\de z'\;.\label{qwl}
}
The source galaxies distribution in the $i$th redshift bin is described as a normalized distribution $n_i(z)$.
In harmonic space, we have the following relation between the convergence field, shear field, and the lensing potential field
\eqsali{
\kappa(\bvec\ell) &=-\dfrac{|\bvec\ell|^2}{2}\phi(\bvec\ell)\;,\\ 
\gamma(\bvec\ell)&=\left(\dfrac{\ell^{2}_{1}-\ell^{2}_{2}+2i\ell_{1}\ell_{2}}{|\boldsymbol{\ell}|^{2}}\right)\kappa(\bvec\ell)=\kappa(\bvec\ell)e^{2i\beta}\;,\label{eq:kgamma}
}
where $\beta$ is the polar angle of $\boldsymbol{\ell}$ \citep{schneider2002b}. 
In galaxy surveys, direct measurement of the convergence field is not possible due to the inability to determine the intrinsic luminosity of source galaxies. However, we can measure the ellipticity of galaxies which contains information about the cosmic shear field $\gamma^G$. Unfortunately, observed ellipticity is subject to contamination from both the galaxy's intrinsic shape and intrinsic alignment induced by its local environment. Consequently, we express the measured shear as $\varepsilon=\varepsilon_s+\gamma^G+\gamma^I$, where $\gamma^I$ represents the intrinsic alignment.
Furthermore, by assuming a white noise statistical property, the intrinsic shape noise power spectrum can be expressed as 
\eqsali{
N^{\varepsilon\varepsilon}_i(\ell)=\dfrac{4\pi f_{\rm sky}}{N_i}\sigma^2_\varepsilon\;,
\label{nggl}
}
where the ellipticity dispersion $\sigma_\varepsilon\approx0.3$ \citep{miao2022cosmological}, $f_{\rm sky}$ is the fraction of sky used in the analysis,  
and $N_i$ is total number of galaxies in the $i$th redshift bin. 

In cosmic shear analysis, in addition to the dominant intrinsic ellipticity of galaxies, there is a second source of contamination, namely intrinsic alignment. This signal can be attributed to its correlation with either the gravitational tidal field of large-scale structures (GI term) or the intrinsic alignment of neighboring galaxies within local environments (II terms). \citep{Troxel:2014dba}.
Hence, the observed angular power spectrum of shear measurement is composed of four components \citep{Bridle_2007}
\eqsali{
\hat C^{\kappa\kappa}_{ij}(\ell)=C^{\kappa\kappa}_{ij}(\ell)+C^{II}_{ij}(\ell)+C^{GI}_{ij}(\ell)+N^{\varepsilon\varepsilon}_i(\ell)\delta_{ij}\;,\label{eq:Cii}
}
where $C^{\kappa\kappa}_{ij}(\ell)$ is the convergence power spectrum, $C^{II}_{ij}(\ell)$ and $C^{GI}_{ij}(\ell)$ are the Intrinsic-Intrinsic (II) and Gravitational-Intrinsic (GI) power and cross spectra, respectively. 

\begin{figure}
    \centering
    \includegraphics[width=\columnwidth]{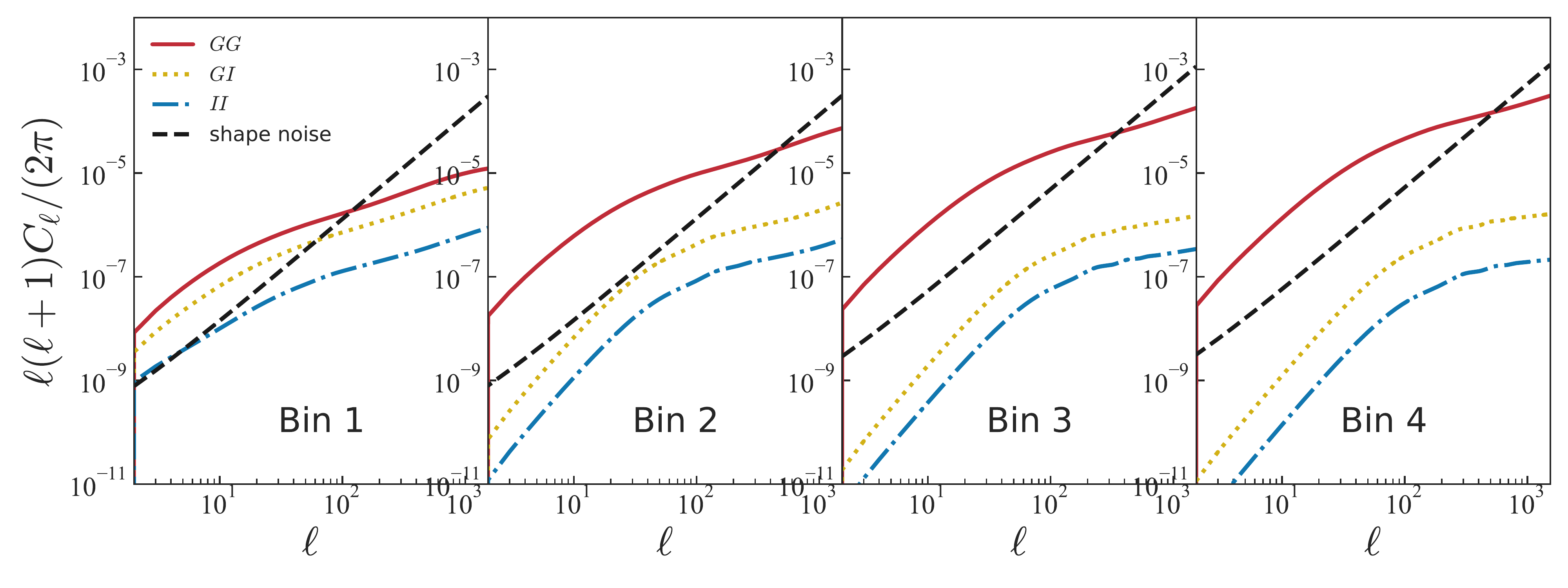}
    \caption{Auto- and cross-spectra of cosmic shear (G), intrinsic alignment (I), and shape noise in 4 photo-$z$ bins.
    The definition of the photo-$z$ bins is described in Fig. \ref{noz}. 
    Red, blue, and yellow curves denote the GG, II, and GI spectra. Black dashed curves denote shape noise. Since the cosmic shear and intrinsic alignment is anti-correlated, 
    to avoid negative values in the log-log plot, we show the absolute value of the GI cross-spectrum.}
    \label{NeGI}
\end{figure}

For simplicity, we adopt the Limber approximation \citep{limber1953analysis} instead of the fast oscillated spherical harmonics, which is approved a good approximation with an error smaller than $2\%$ on the scales of $\ell>20$ \citep{Kilbinger:2017lvu}
\eqsali{
C^{\kappa\kappa}_{ij}(\ell)=\dint_0^{z_*}\dfrac{c\de z}{H(z)}\dfrac{q_i(z)q_j(z)}{\chi^2}P_{\delta}(k=\dfrac{\ell+1/2}{\chi},z)\;,
\label{eq:conv1}
}
The Intrinsic-Intrinsic and Gravitational-Intrinsic power and cross spectra are given assuming the non-linear linear alignment (NLA) model \citep{Hirata:2004gc,Bridle_2007}
\eqsali{
C^{II}_{ij}(\ell)=\dint_0^{z_*}\dfrac{c\de z}{H(z)}\dfrac{n_i(z)n_j(z)f_i(z)f_j(z)}{\chi^2}P_{\delta}(k=\dfrac{\ell+1/2}{\chi},z)\;,
\label{eq:iihu}
}
and
\eqsali{
C^{GI}_{ij}(\ell)=\dint_0^{z_*}\dfrac{c\de z}{H(z)}\dfrac{n_i(z)f_i(z)q_j(z)}{\chi^2}P_{\delta}(k=\dfrac{\ell+1/2}{\chi},z)+
\dint_0^{z_*}\dfrac{c\de z}{H(z)}\dfrac{n_j(z)f_j(z)q_i(z)}{\chi^2}P_{\delta}(k=\dfrac{\ell+1/2}{\chi},z)\;,
\label{eq:conv2}
}
where $f_i(z)$ is written as
\eqsali{
f_i(z)=-A_{\rm IA}C_1 \rho_{\rm crit} \dfrac{\Omega_m}{D(z)}\left(\dfrac{1+z}{1+z_0}\right)^{\eta_{\rm IA}}\left(\dfrac{L_i}{L_0}\right)^{\beta_{\rm IA}}\;, \label{eq NLA IA}
}
and $C_1=5\times 10^{-14} h^{-2}M_{\odot}^{-1}{\rm Mpc}^{3}$ as in \cite{Bridle_2007}, $\rho_{\rm crit}$ is the present critical density, $D(z)$ is the linear growth factor normalized to unity at $z=0$, and $z_0=0.6$ and $L_0$ are pivot redshift and luminosity, respectively. Since the change of average luminosity can be ignored, we don't consider luminosity dependence and the fiducial values of $A_{\rm IA}$, $\eta_{\rm IA}$ and $\beta_{\rm IA}$ are set to be 1, 0, 0, respectively \citep{joudaki2016cfhtlens}.
In Fig. \ref{NeGI}, we show the auto- and cross-spectra of the cosmic shear and intrinsic alignment in 4 photo-$z$ bins, which are defined in the following sections. First of all, the intrinsic alignment model we adopted is anti-correlated with cosmic shear. Second, the amplitude of the II correlation is much smaller than GG. Taking these two effects together into account, we can conclude that the $\hat C^{\kappa\kappa}_{\ell}$ in the Eq. (\ref{eq:Cii}) gets smaller once we consider the intrinsic alignment. 

\subsection{Weak lensing-CMB lensing cross-correlation}

Since the convergence field and shear field are both determined by the gravitational potential $\phi$, the CMB lensing convergence field and the cosmic shear are correlated. Furthermore, we can find a linear combinations of $\gamma_1$ and $\gamma_2$ to convert shear field into E and B mode ($\gamma_E$, $\gamma_\mathrm{B}$)
\eqsali{
\gamma_E = \gamma_1(\bvec\ell)\dfrac{\ell^2_1-\ell^2_2}{\ell^2_1+\ell^2_2}+\gamma_2(\bvec\ell)\dfrac{2\ell_1\ell_2}{\ell^2_1+\ell^2_2}=\kappa(\bvec\ell);\quad \gamma_\mathrm{B}=0\;.\label{spin-2}
}
The cross-spectrum of weak lensing and CMB lensing field  $C^{\kappa_{\rm CMB}\gamma_E}_i(\ell)$ reads
\eqsali{
\langle\kappa_{\rm CMB}(\bvec\ell)\gamma^*_\mathrm{E,i}(\bvec\ell')\rangle=(2\pi)^2\delta_D(\bvec\ell-\bvec\ell')C^{\kappa_{\rm CMB}\gamma_E}_i(\ell)\;,\label{comcl}
}
where $\delta_D$ is the 2-dimensional Dirac function and the sub-index $i$ denotes for the $i$th redshift bin. Under the Limber approximation, the theoretical cross-spectrum reads 
\eqsali{
C^{\kappa_{\rm CMB}\gamma_E}_i(\ell)=\dint_0^{z_*}\dfrac{c\de z}{H(z)}\dfrac{q_\mathrm{CMB}(z)q_i(z)}{\chi^2}P_{\delta}(k=\dfrac{\ell+1/2}{\chi},z)\;,
}
where $P_\delta(k,z)$ is the nonlinear matter power spectrum, which is calculated with \verb'pyccl' \citep{Chisari_2019} and with the \verb'HALOFIT' method for describing the nonlinear part \citep{smith2003stable,takahashi2012revising}. 
The cross-correlations are also affected by the IA according to Eq. (\ref{eq:conv2}). As shown in Ref. \cite{Troxel:2014dba,hall2014intrinsic}, IA will suppress CMB Lensing-cosmic shear cross spectrum by roughly $15\%$.

\begin{figure}
    \centering
    \includegraphics[width=0.6\columnwidth]{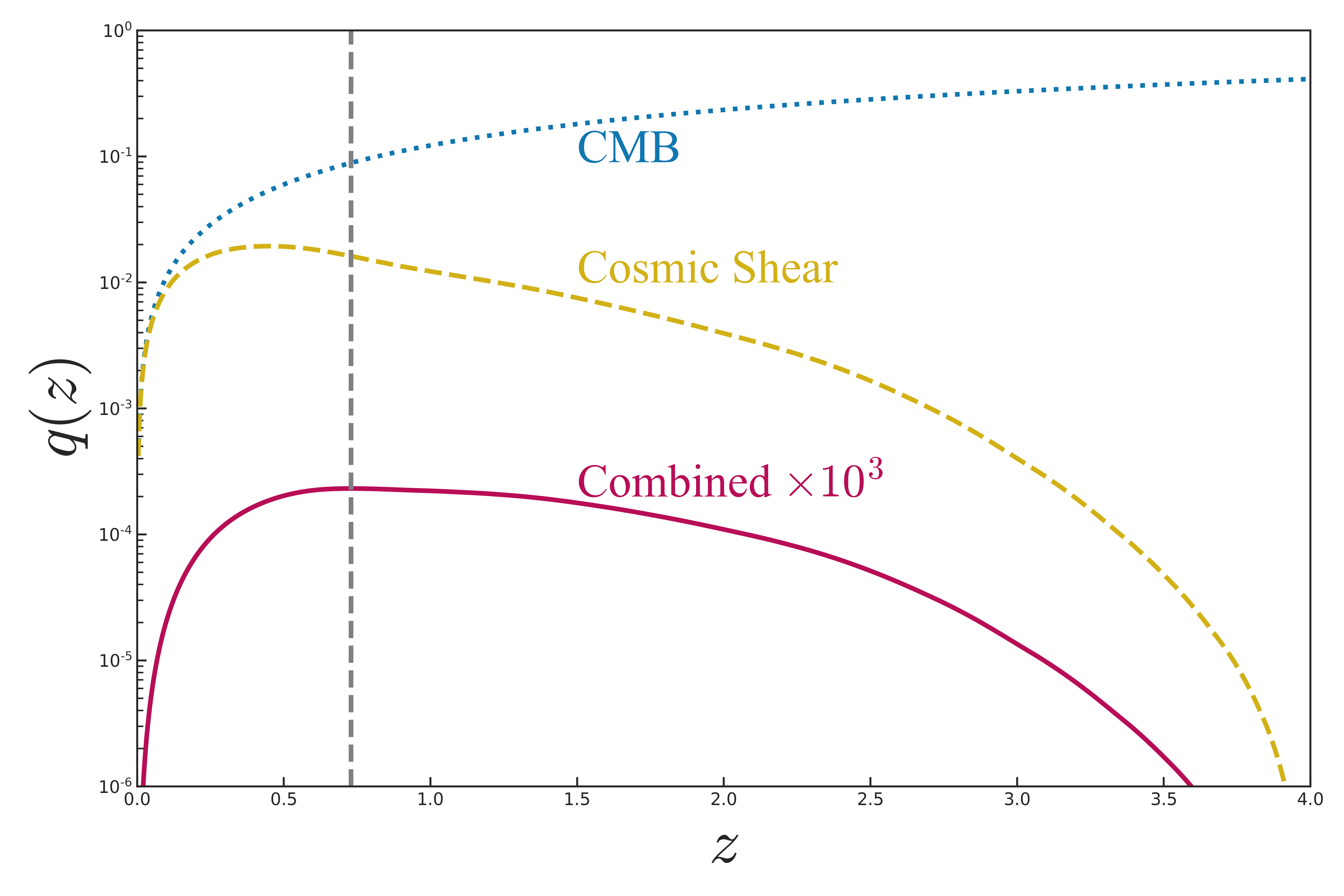}
    \caption{Lensing efficiencies that enter the cosmic shear-CMB lensing correlations as defined in Eq. (\ref{qcmb}) and Eq. (\ref{qwl}). The vertical gray line marks the effective redshift $z\approx0.73$ for the cosmic shear-CMB lensing cross-correlation signal, measured from the nominal CSST number density. For cosmic shear, we used the CSST source sample redshift distribution shown in Fig. \ref{noz}. The combined lensing efficiency is defined in Eq. (\ref{eq:combine}). 
}
    \label{kernels}
\end{figure}

\begin{figure}
    \centering
    \includegraphics[width=0.6\columnwidth]{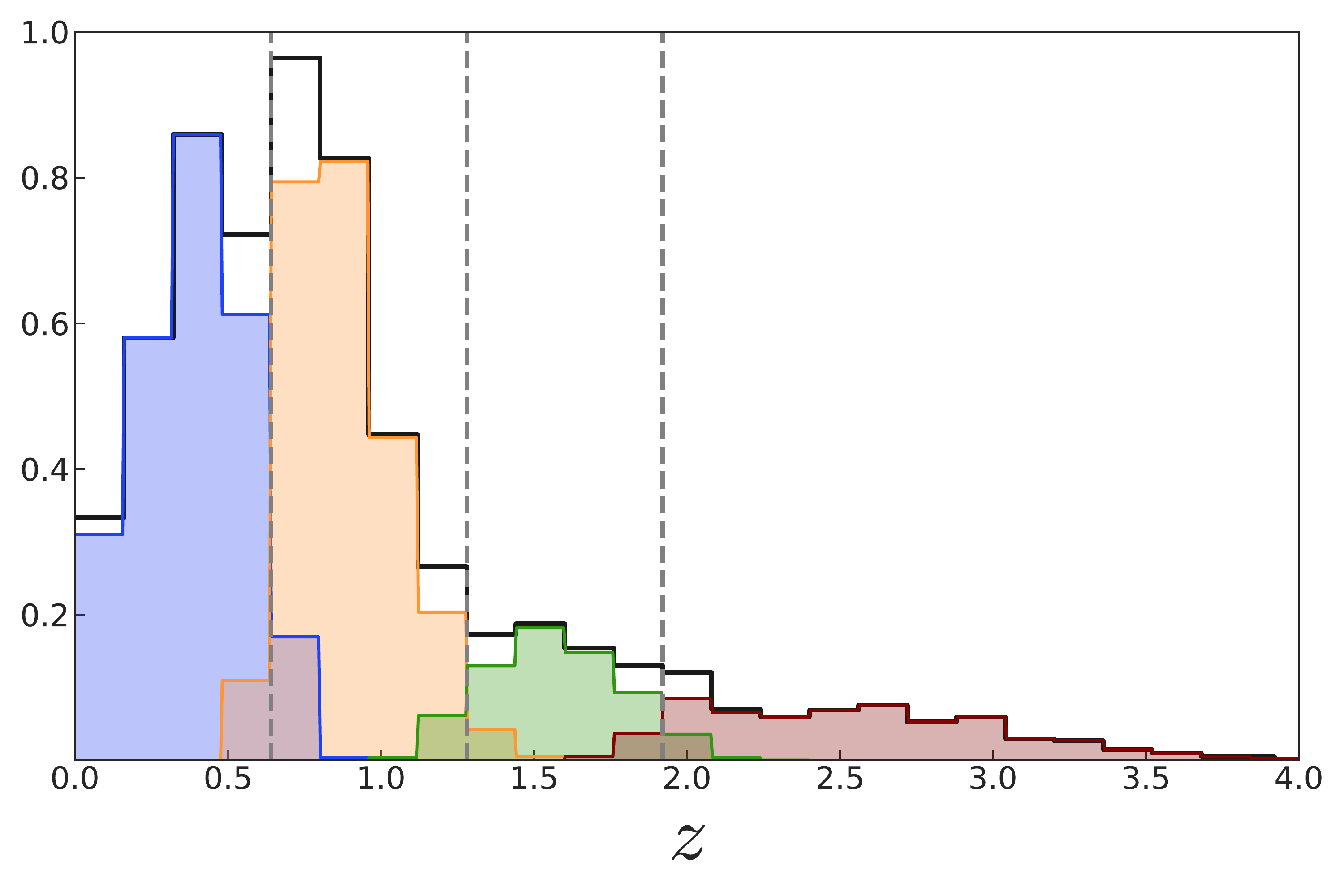}
    \caption{The mock galaxy redshift distribution in the photometric imaging survey of the CSST. The black line denotes the total redshift distribution $n(z)$, which is obtained from the COSMOS-2015 catalog \citep{2018MNRAS.480.2178C}. The blue, orange, green, and red curves are the true redshift distribution $n_i(z)$ in each of the four photo-$z$ bins. The four photo-$z$ bins are divided by the gray vertical dashed curves.}
    \label{noz}
\end{figure}

In Fig. \ref{kernels}, we show the lensing efficiency of CMB lensing
and cosmic shear\footnote{calculated using the CSST source mock
redshift distribution (solid black curve in Fig.\ref{noz})} as a function of redshift, as well as the combined weight given by
\eqsali{
q^{\rm combined}(z)=\dfrac{H(z)}{c}D^2(z)q_{\rm CMB}(z)q_{\gamma}(z)\;,
\label{eq:combine}
}
where $D(z)$ is the growth function normalized to the unit at $z=0$, accounts for the growth of matter perturbations with redshifts, and $q_{\gamma}(z)$ is lensing efficiency of cosmic shear with the source distribution in the entire range of redshift $n(z)$, shown in Fig. \ref{noz}.

\section{Simulation}\label{sec3}

In this section, we present our simulations to generate correlated signals between CMB lensing and cosmic shear. The sky coverage of the CSST wide field is 17,000 deg$^2$ \citep{zhanhu}, while the AliCPT-1 deep patch's sky coverage is approximately 4,000 deg$^2$ in the high latitude of the northern hemisphere \citep{liu2022forecasts}. Therefore, the entire AliCPT-1 patch is covered by CSST. In our signal generation, we adopt the flat sky approximation for simplicity. To match this approximation, we choose a 30$\times$30 deg$^2$ sky area for the cross-correlation computation. However, it should be noted that considering only a sky overlap of 900 deg$^2$ instead of the full 4000 deg$^2$ is a significant limitation of this work. While using the full field of AliCPT-1 would yield more realistic results, it would also require a more complicated spherical harmonics transformation than the simple Fourier transformation used in the flat field. As such, we have adopted the flat-sky approximation in this paper. To the best of our knowledge, this approximation becomes increasingly inaccurate for $\ell<10$ \citep{Lemos:2017arq}, which corresponds to an angular scale greater than roughly 30 degrees.

In order to generate the correlated signals between CMB lensing and cosmic shear, the most straightforward way is to measure these two quantities in the same mock catalog. 
However, due to the unmatched lensing kernel between WL and CMB lensing, it will be very expensive to do a large sky coverage ray tracing through N-body simulation until the redshift range where CMB lensing efficiency is enough high. To evade this obstacle, we generate the correlated cosmic shear and CMB lensing signals from the Gaussian realizations based on the inputted auto- and cross-spectra. We generate both CMB lensing $\kappa$ map and cosmic shear with \verb'HEALPix'\citep{2005ApJ...622..759G,Zonca2019}. The resolution is chosen as $N_{\rm side}=512$, corresponding to a pixel size of about $7'$. 
We choose the same Gaussian window function for both the CMB lensing convergence and WL shear maps, which equals $\ell_*=800$, $W_\ell=\exp[-\ell^2/\ell_*^2]$. We neglect the baryonic effect since it's only sensitive on scales smaller than $\ell=1500$ \citep{joachimi2021kids}. To avoid the leakage from the sharp edges due to the Fourier transformation, we apodize the maps with a cosine function and apodization scale of $2^\circ$.\footnote{In \texttt{NaMaster} algorithm, the mode-coupling due to the partial sky effect is mitigated by inverting the mode-coupling matrix in the calculation of the pseudo-$C_{\ell}$. Hence, the apodization is not a necessary step anymore. We have verified that whether or not to perform apodization on the mask does not affect the measurement of pseudo-$C_\ell$. We appreciate the anonymous referee for pointing this out.}
In order to generate a set of maps that satisfy the desired auto- and cross-spectrum of CMB lensing and cosmic shear, we extend Kamionkowski's method \citep{kamionkowski1997statistics} to an arbitrary number of maps\footnote{A similar expression is also derived in \cite{Giannantonio:2008zi}}. The maps we are going to simulate in harmonic space are written as
\eqsali{\label{eq:mmap}
M_i(\bvec{\ell})=\zeta_0(\bvec{\ell})s_{0i}+\zeta_1(\bvec{\ell})s_{1i}+...+\zeta_i(\bvec{\ell})s_{ii}\;(i=0,1,2,...,n)
}
$s$ has the following recursive form
\eqsali{s_{ij}=\dfrac{C_{ij}-\dsum^{i-1}_{k=0}s_{ki}s_{kj}}{\displaystyle\left(C_{ii}-\dsum^{i-1}_{k=0}s^2_{ki}\right)^{1/2}}\;,\label{eq build map}}
where $\zeta_{i}(\ell)$s are two independent complex numbers drawn from a Gaussian distribution with zero mean and unit variance. $C_{ij}$ denotes the auto- and cross-spectrum, for $i,j=0$ we have $s_{00}=\sqrt{C_{00}}$.
We would like to emphasize that the method we presented here is not novel. There are pioneer works that have already implemented this algorithm into CMB (including integrated Sachs–Wolfe effect) and galaxy number count cross-correlations, such as \cite{Giannantonio:2008zi}. Besides, several commonly used software in the cosmology community have also implemented this algorithm, such as \verb'healpy' \citep{Zonca:2019vzt}, \verb'HEALPix' \citep{2005ApJ...622..759G} and \verb'NaMaster' \citep{10.1093/mnras/stz093}.

\subsection{CMB lensing map}

\begin{figure}
    \centering
    \includegraphics[width=0.6\columnwidth]{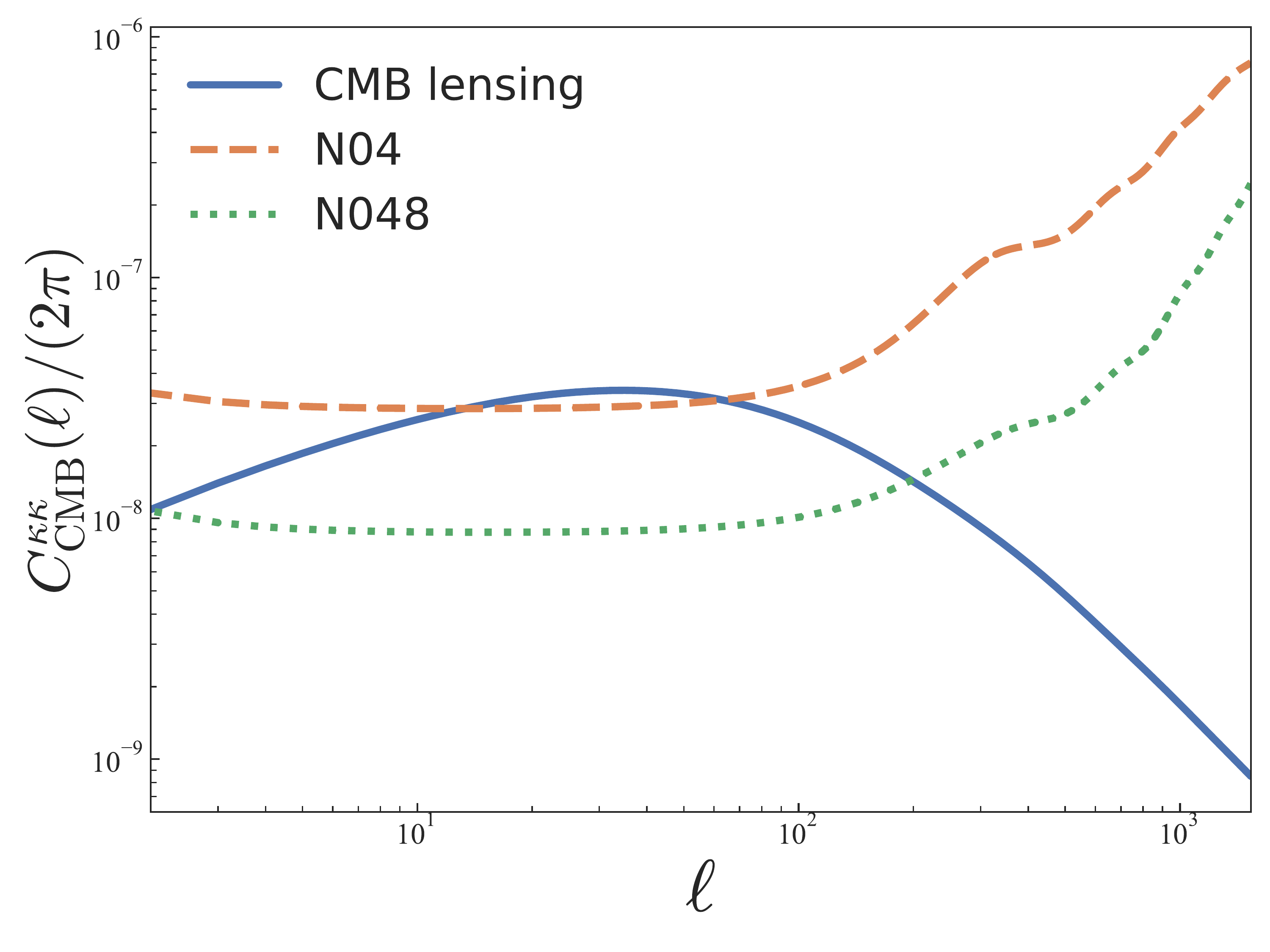}
    \caption{The angular power spectrum of CMB lensing and reconstruction noise of AliCPT-1. The blue curve represents the theoretical angular power spectrum of CMB lensing convergence, while the dashed orange and green dotted curves depict the reconstruction noise angular power spectrum from``4 module*yr'' and ``48 module*yr''.}
    \label{Nphi}
\end{figure}

In CMB lensing, we generate the convergence signal spectrum using \verb'pyccl', while the noise spectrum is obtained from AliCPT mocks using both temperature and polarization \citep{liu2022forecasts, disandai}
Our analysis employs an observed patch that covers about 12\% of the sky \citep{li2018tibet}. We investigate two noise levels in this study, which are depicted by the orange dashed and green dotted curves in Fig.\ref{Nphi} for ``4 module*yr'' and ``48 module*yr'', respectively. For simplicity, we only consider the leading N0 noise and ignore sub-leading contributions like N1 noise, etc. We obtain the harmonic coefficients of the convergence map by adopting $C_{00}=C^{\kappa\kappa}_{\rm CMB}(\ell)$ and $C_{0i}=C^{\kappa_{\rm CMB}\gamma_E}_i(\ell)$\\
and the noise map is defined as
\eqsali{n_\kappa(\boldsymbol{\ell})=\zeta'(\boldsymbol{\ell})(N^{\kappa\kappa}_{\rm CMB}(\ell))^{1/2}\;,}
$C^{\kappa\kappa}_{\rm CMB}(\ell)$ is theoretical convergence power spectrum and $N^{\kappa\kappa}_{\rm CMB}(\ell)$ is noise power spectrum. 
The noise spectrum is directly extracted from the AliCPT-1 lensing reconstruction outcomes \citep{liu2022forecasts}.
In Fig.\ref{k_g_maps}, the first row shows the map of CMB lensing convergence. The simulated CMB lensing convergence map in harmonic space is given by $\hat\kappa_{\rm CMB}(\boldsymbol{\ell})=\kappa_{\rm CMB}\boldsymbol{\ell})+n_\kappa(\boldsymbol{\ell})$.

\subsection{Galaxy samples and cosmic shear}

In this section, we describe the methodology used to generate our cosmic shear maps. To create the simulated convergence and shear maps, we utilized Eq. (\ref{eq:conv1})-(\ref{eq:conv2}). The IA signals were implemented in the power spectra following Eq.,\eqref{eq:Cii}. To validate these maps, we compared the reconstructed auto-spectrum with the theoretical ones. We considered three types of errors or biases: photo-$z$ error, galaxy shape error, and intrinsic alignment. In this work, we assumed that these three errors are independent. While we did not model the intrinsic alignment at the map level, we treated it as part of the shear signal and modeled it in the angular power spectrum in Eq. (\ref{eq:Cii}). Two types of shear power spectrum models were considered: one with only cosmic shear signals (model I) and another with both cosmic shear signals and intrinsic alignment (model II). As previously demonstrated, since there is an anti-correlation between cosmic shear and intrinsic alignment, the amplitude in model II is smaller than in model I. 
The total galaxy number density is assumed to be $20~{\rm arcmin}^{-2}$. We divide the source galaxy samples into 4 bins as it is shown.
We use this galaxy redshift distribution $n_i(z)$ to calculate the cosmic shear signal according to Eq. (\ref{qwl}). We generate the harmonic coefficients of the convergence map according to Eq. (\ref{eq:mmap}) with $C_{ij}=C^{\gamma_{E}\gamma_{E}}_{ij}(\boldsymbol{\ell})$ 
\eqsali{
\kappa^{sig}_i(\boldsymbol{\ell})=\dsum^i_k\zeta_k(\boldsymbol{\ell})s_{ki}\;.\label{eq shear map}
}
Here, we emphases that Eq. (\ref{eq:mmap}) is only applicable for the spin-0 field, namely the convergence field. To get the shear field (spin-2 field), we need to go through Eq. (\ref{eq:kgamma}), where $\beta$ is the polar angle of the vector $\bvec\ell$.
The shape noise is generated according to its power spectrum given by Eq. (\ref{nggl}). The noise level is shown in Fig. \ref{NeGI} and is jointly determined by the galaxy number density ($n_i$) and the ellipticity dispersion of a single galaxy ($\sigma_\varepsilon$).  

\begin{figure*}
    \centering
    \includegraphics[width=\columnwidth]{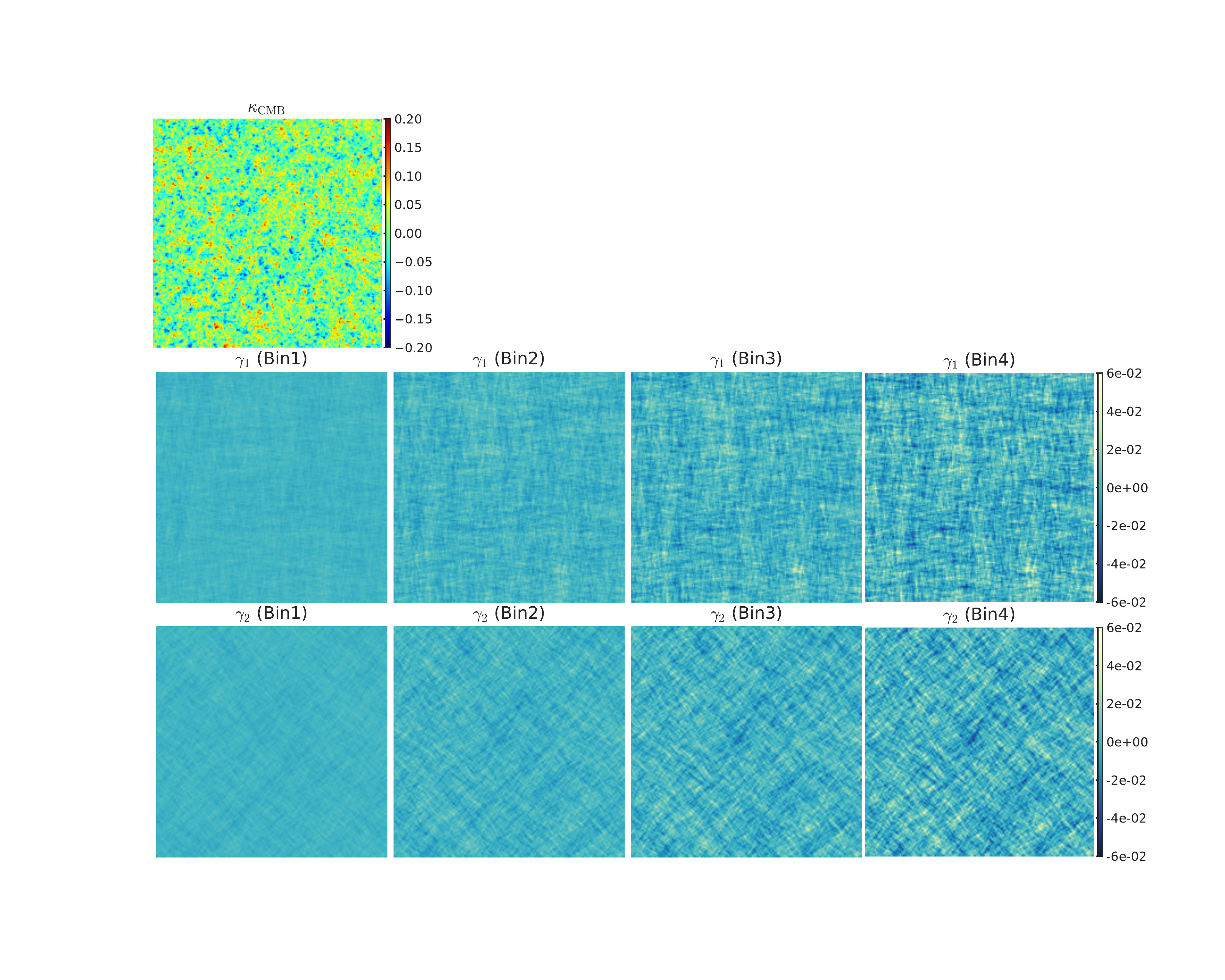}
    \caption{CMB lensing convergence map and cosmic shear maps in 4 redshift bins. The map in the first row is the CMB convergence. The second and third rows, from left to right, are the real and imaginary parts of the plural cosmic shear from the first to the fourth redshift bins, respectively.}
    \label{k_g_maps}
\end{figure*}

Due to the errors in determinating photometric redshift, the real galaxy redshift distribution in the $i$th photo-$z$ bin are conventionally expressed as eg. \citep{ma2006effects}
\eqsali{
n_i(z)=\dint^{z^P_{i,{\rm max}}}_{z^P_{i,{\rm min}}}\de z^P n(z) p(z^P|z)\;,
}
where $z^p$ is photo-$z$, $n(z)$ is the total redshift distribution (shown in Fig. \ref{noz}), which is calculated from COSMOS-2015 catalog  
by applying the observational limits from the CSST bands \citep{Liu2023,2018MNRAS.480.2178C}.
$p(z^P|z)$ is the photo-$z$ distribution function given the real redshift $z$
\eqsali{
p(z^P|z)=\dfrac{1}{\sqrt{2\pi }\sigma_z(1+z)}\exp\left[-\dfrac{(z-z^P-\Delta^i_z)^2}{2(\sigma_z(1+z))^2}\right]\;,\label{photo-$z$-eq}
}
where $\Delta_z$ and $\sigma_z$ are the redshift bias and scatter, respectively. In this work, we adopt the typical values in the 4th generation surveys as $\Delta_z=0.005$ and $\sigma_z=0.05$. 
The photo-$z$ errors in the maps arise via redistributing the true galaxy redshift according to the above probability distribution. 
For the pixelized representation of cosmic shear catalogs, we construct re-weighted tomographic maps as
\eqsali{
\hat{\gamma}_i(p)=\dfrac{\dsum_{j}w_{j\rightarrow i}(p)(\gamma^{\rm sig}_j(p)+\varepsilon_j(p))}{\dsum_{j}w_{j\rightarrow i}(p)}\;,
}
where $p$ denotes pixel index, $w_j$ are the probability of a shear signal leaking from $j$th redshift bin to $i$th photo-$z$ bin. The weights are drawn by multigaussian distribution $w_{j\rightarrow i}\sim M\{N_{bins},[p_{j\rightarrow i}]\}$
\eqsali{
p_{j\rightarrow i}=\dint^{z^P_{i,{\rm max}}}_{z^P_{i,{\rm min}}}\de z^P p(z^P|z_j)=\dfrac{1}{2}\left[{\rm erf}{(x_{j\rightarrow i}^{\rm max})}-{\rm erf}{(x_{j\rightarrow i}^{\rm min})}\right]\;,\label{leakex}
}
with
\eqsali{
x_{j\rightarrow i}^{\rm max/min}=\left[\dfrac{(z_j-z^P_{i,{\rm max/min}}-\Delta^j_z)^2}{2(\sigma_z(1+z_j))^2}\right]\;,
}
where ${\rm erf}(x)$ is error function.
The resulting shear maps are shown in the second row and third row in Fig. \ref{k_g_maps}.

\section{pseudo power spectra measurement}\label{sec4}

We generated 600 simulated maps of the CMB convergence field and the cosmic shear field using the same method but with different random seeds. The angular cross spectra between the CMB and shear maps were computed using a pseudo-$C_\ell$ estimator based on \verb'NaMaster' algorithm \citep{10.1093/mnras/stz093}employing Eq. (\ref{comcl}). 
To accelerate the covariance computation and invert the mode coupling matrix in the pseudo-$C_\ell$ calculation, we binned the cross-spectrum in $\ell$.
We utilized 19 linearly spaced multipole bins with a width of $\Delta\ell=40$ in the range $20\le\ell\le800$, resulting in estimated band powers denoted as $\hat{C}^{\kappa_{\rm CMB}\gamma_{E}}_i(L)$, where $L$ denotes the multipole bin and $i$ for the photo-$z$ bin.
To simplify the index notations, we merged the multipole index $L$ and photo-$z$ bin index $i$ into a single index $\nu=[L^{(1)}, L^{(2)}, L^{(3)}, L^{(4)}]$, where we sequentially list the banded multipoles in each photo-$z$ bin. In order to highlight the higher multipoles visually, we use the multipole-weighted spectrum 
$D_\nu=L C_\nu B_L^{-1}$, where $C_\nu=[\hat{C}^{\kappa_{\rm CMB}\gamma_{E}}_1(L),\hat{C}^{\kappa_{\rm CMB}\gamma_{E}}_2(L),\hat{C}^{\kappa_{\rm CMB}\gamma_{E}}_3(L),\hat{C}^{\kappa_{\rm CMB}\gamma_{E}}_4(L)]$ and $B_L$ is the beam function in the band power. In the spectrum estimation, we de-convolved the beam. 
Finally, we de-convolved the beam in the spectrum estimation to obtain the covariance matrix, which reads as follows:
\eqsali{
\mathbb{C}_{\nu\nu'}=\dfrac{1}{N_{\rm sim}-1}\dsum^{N_{\rm sim}}_{\alpha=1}[\hat{D}^{XY,\alpha}_\nu-\bar{D}^{XY}_\nu][\hat{D}^{XY,\alpha}_{\nu'}-\bar{D}^{XY}_{\nu'}]\;,
}
with $X,Y=\{\kappa_{\rm CMB},\gamma_E\}$ and $\alpha$ denotes the number of the simulation. 
$\hat{D}^{XY,\alpha}_\nu$ is the estimated cross spectrum from the $\alpha$th simulation and $\bar{D}^{XY}_\nu$ is the mean over 600 simulations
\eqsali{
\bar{D}^{XY}_\nu=\dsum^{N_{\rm sim}}_{\alpha=1}\dfrac{\bar{D}^{XY,\alpha}_\nu}{N_{\rm sim}}\;.
}
To calculate the inverse covariance, we adopt the unbiased
estimator used in \citealt{2007A&A...464..399H}, which is given by
\eqsali{\hat{\mathbb{C}}^{-1}_{\nu\nu'}=\dfrac{N_{\rm sim}-N_{\rm bin}-2}{N_{\rm sim}-1}\mathbb{C}_{\nu\nu'}^{-1}\;,}
where $N_{\rm sim}=600$ and $N_{\rm bin}=4\times19$ is the number of data points and $\mathbb{C}_{\nu\nu'}^{-1}$ is
the normal inverse of $\mathbb{C}_{\nu\nu'}$. In order to include the error propagation from the error in the covariance matrix into the fitting parameters \citep{percival2014clustering} we rescale the covariance matrix, 
\eqsali{
\Tilde{\mathbb{C}}_{\nu\nu'}^{-1}=\dfrac{1+B(N_{\rm bin}-N_{\rm p})}{1+A+B(N_{\rm p}+1)}\hat{\mathbb{C}}^{-1}_{\nu\nu'}
}
here $N_{\rm p}$ is the number of the fitting parameters, and
\eqsali{
A=\dfrac{2}{(N_{\rm sim}-N_{\rm bin}-1)(N_{\rm sim}-N_{\rm bin}-4)}
}
\eqsali{
B=\dfrac{N_{\rm sim}-N_{\rm bin}-2}{(N_{\rm sim}-N_{\rm bin}-1)(N_{\rm sim}-N_{\rm bin}-4)}
}
\begin{figure*}
    \centering
    \includegraphics[width=\columnwidth]{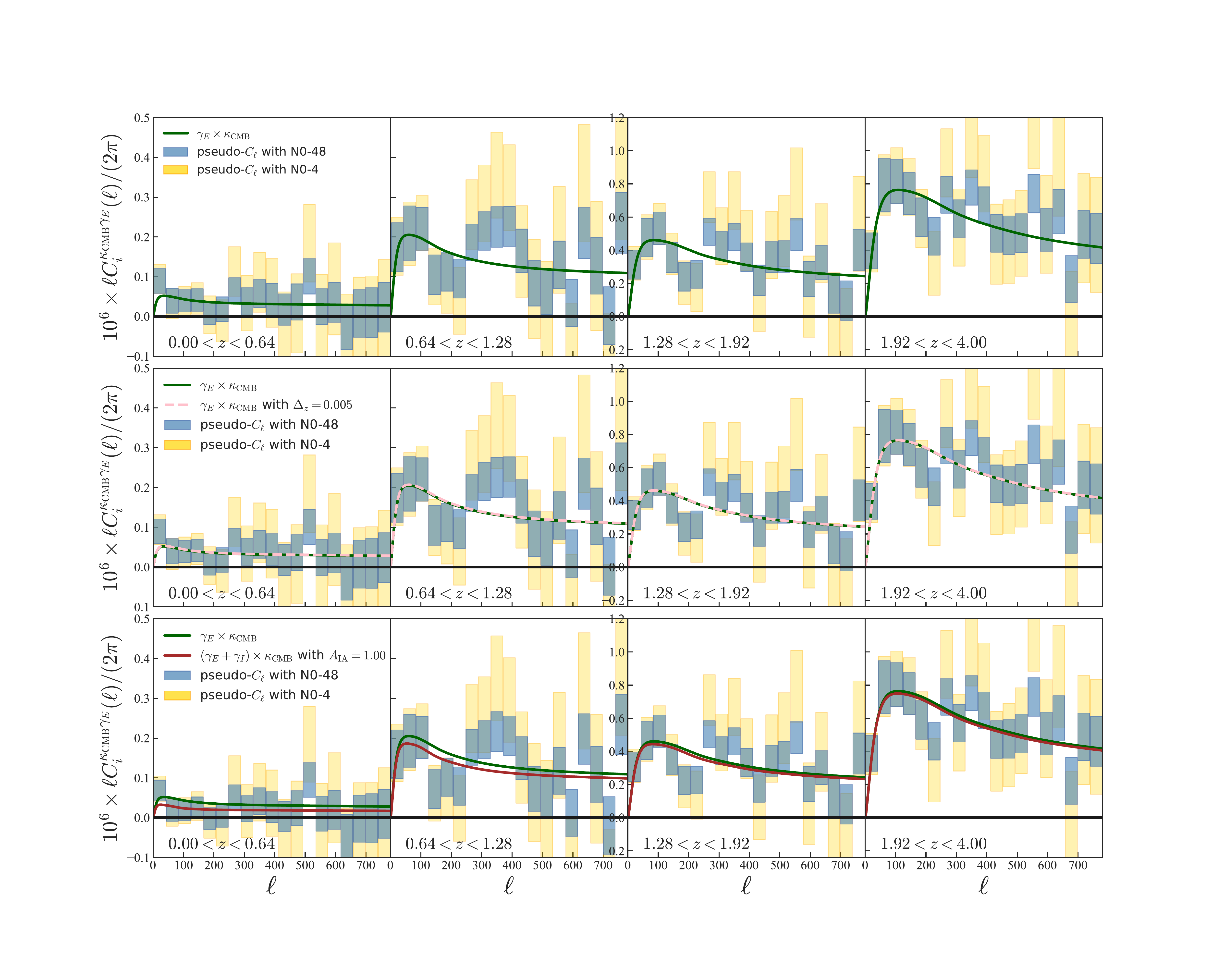}
    \caption{The estimation of the shear-CMB cross-spectrum in four photo-$z$ bins. The yellow and blue boxes are the binned error bars in multipoles, with two different CMB lensing reconstruction N0 noises. The first row is the pseudo-$C_\ell$ estimation without any biases. The second row is the one with redshift bias  $\Delta_z=0.005$. In the third row are the results with intrinsic alignment amplitude $A_{\rm IA}=1$. The solid curves are the theoretical predictions. The spectrum covariances are estimated from 600 mocks with N0-4 and N0-48 noise setups.}
    \label{pseudo-Cls}
\end{figure*}

\begin{figure}
    \centering
    \includegraphics[width=0.8\columnwidth]{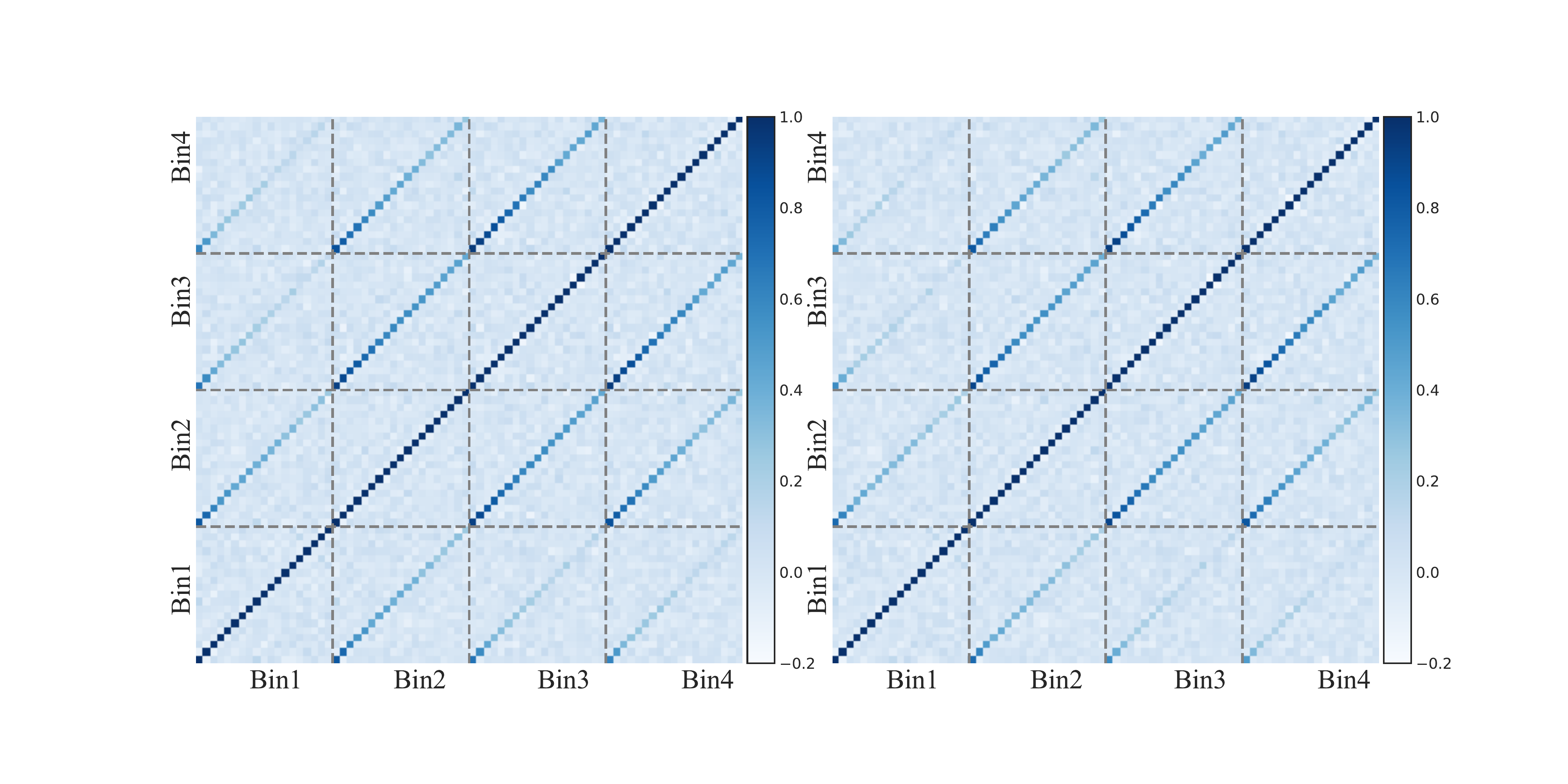}
    \caption{Normalized covariance matrix of multipole and photo-$z$ bins of cross-spectra. The left and right panels are for the N0-4 and N0-48 CMB lensing noise cases. }
    \label{N048-coe}
\end{figure}

We present Fig. \ref{pseudo-Cls}, which displays Pseudo-$C_\ell$s associated with different noise types. The yellow and blue boxes represent the error bars originating from AliCPT-1's  ``4 modules*yr'' and ``48 modules*yr'' experimental setups, respectively, whose noise spectra correspond to the orange dashed and green dotted curves in Fig. \ref{Nphi}. The first row in Fig. \ref{pseudo-Cls} is calculated without any biases, while the green solid curve represents the theoretical cross-spectrum. In the second row, a photo-$z$ bias $\Delta_z=0.005$ is added. By visual inspection, the red solid curve (with the bias) and the green solid curve (without the bias) appear indistinguishable. In the third row, intrinsic alignment effects are incorporated into the simulations, where the green solid curve still represents the theoretical shear-CMB cross-correlation, while the red solid curve assumes an intrinsic alignment amplitude $A_{\rm IA}=1$. Notably, the intrinsic alignment can significantly reduce the signal at lower redshifts.

\begin{table}
    \centering
\begin{tabular}{lccccc}
\hline
 S/N\quad & Bin1  & Bin2  & Bin3  & Bin4  & Total \\
\hline
N04               &$ 6.35$&$ 7.83$&$ 10.50$&$ 12.91$&$ 15.01$\\
N04 with bias-$z$ &$ 6.38$&$ 7.87$&$ 10.53$&$ 12.94$&$ 15.04$\\
N04 with IA       &$ 5.60$&$ 7.00$&$ 10.08$&$ 12.65$&$ 14.70$\\
N048              &$ 7.42$&$ 11.53$&$ 16.00$&$ 21.15$&$ 22.63$\\
N048 with bias-$z$&$ 7.49$&$ 11.60$&$ 16.05$&$ 21.16$&$ 22.66$\\
N048 with IA      &$ 6.21$&$ 10.26$&$ 15.41$&$ 20.77$&$ 22.20$\\
\hline
\end{tabular}
    \caption{Signal-to-noise ratio of pseudo-$C_\ell$s measured from 4 photo-$z$ bins and the combined one. Different CMB lensing noise, galaxy photo-$z$ bias as well as intrinsic alignment are considered.}
    \label{snr}
\end{table}

\begin{table}
    \centering
\begin{tabular}{lccccc}
\hline
$\chi^2/N_{\rm bin}$\quad & Bin1  & Bin2  & Bin3  & Bin4  & Total \\
\hline
N04               &$ 1.20$&$ 1.10$&$ 1.24$&$ 1.19$&$ 1.06$\\
N04 with bias-$z$ &$ 1.20$&$ 1.10$&$ 1.24$&$ 1.19$&$ 1.06$\\
N04 with IA/G     &$ 1.00$&$ 1.26$&$ 1.31$&$ 1.28$&$ 1.03$\\
N04 with IA/G+I   &$ 1.11$&$ 1.12$&$ 1.25$&$ 1.23$&$ 1.05$\\
N048              &$ 1.18$&$ 1.09$&$ 1.37$&$ 1.30$&$ 1.09$\\
N048 with bias-$z$&$ 1.18$&$ 1.09$&$ 1.37$&$ 1.30$&$ 1.09$\\
N048 with IA/G    &$ 1.02$&$ 1.42$&$ 1.54$&$ 1.43$&$ 1.11$\\
N048 with IA/G+I  &$ 1.10$&$ 1.14$&$ 1.39$&$ 1.33$&$ 1.09$\\
\hline
\end{tabular}
    \caption{Reduced $\chi^2$ measured from 4 photo-$z$ bins and the combined one. The name "with IA/G+I" denotes the case that both in the simulation and spectrum estimation procedures, the intrinsic alignment is consistently concluded. The one "with IA/G" denotes the case that only in the simulation step the intrinsic alignment is included but not in the step of spectrum estimation. Here $N_{\rm bin}=4\times19$ is our total data number and $N_{\rm bin}=19$ for each bin. }
    \label{rechi2}
\end{table}

\begin{figure}
    \centering
    \includegraphics[width=450pt]{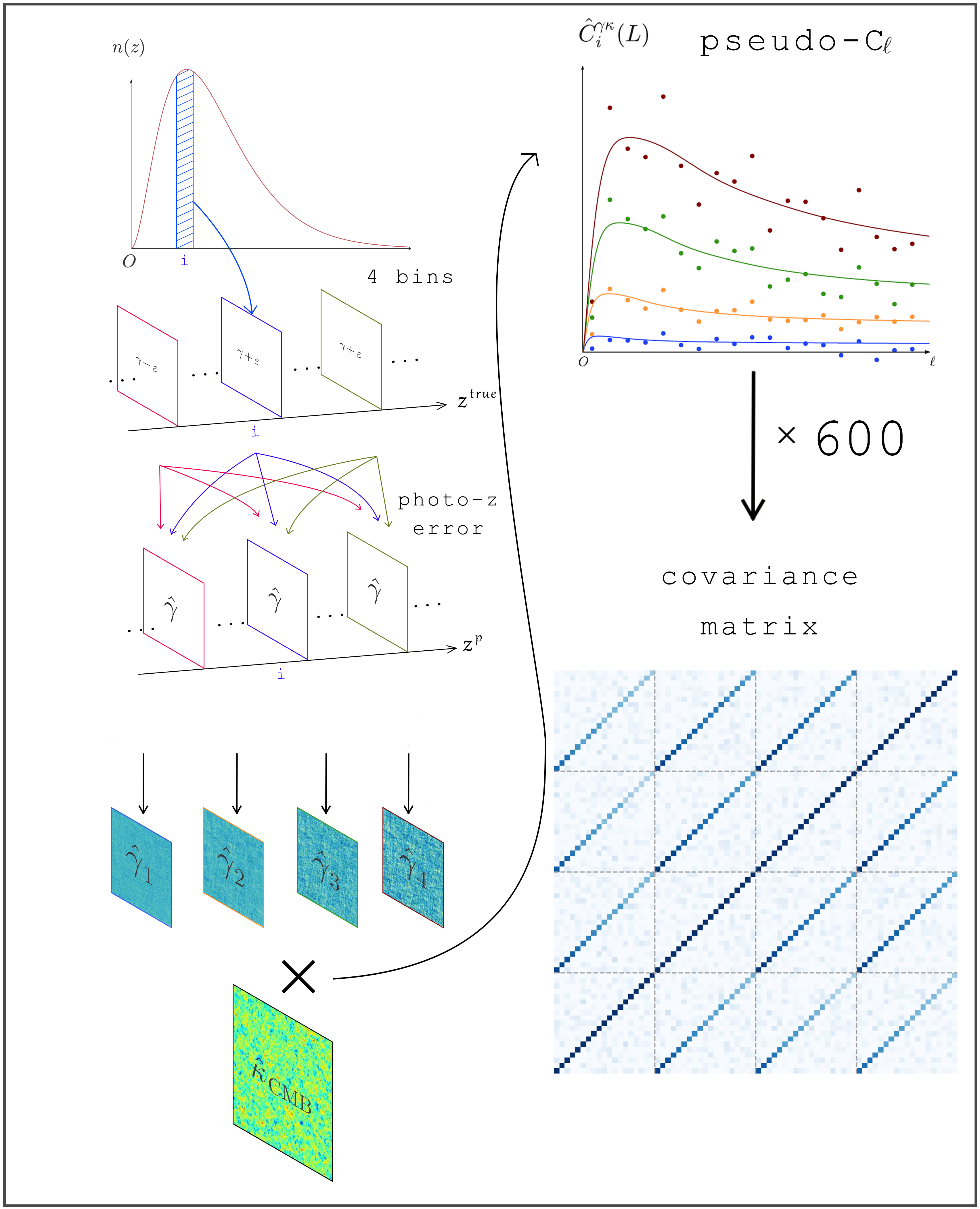}
    \caption{Sketch picture of photo-$z$ error, pseudo-$C_\ell$ measuerent and covariance matrix computing method.}
    \label{ppl}
\end{figure}

We report the signal-to-noise ratio (SNR), which is computed as
\eqsali{
{\rm SNR}=\sqrt{\dsum_{\nu\nu'}\hat{D}^{XY}_\nu\hat{\mathbb{C}}^{-1}_{\nu\nu'}\hat{D}^{XY}_{\nu'}}\;,
}
where $\hat{D}^{XY}_\nu$ is generated by a new simulation that is independent of the other 600 ones. The normalized covariance matrices of cross-correlation are shown in Fig. \ref{N048-coe}. One can see that the correlations of the same multipoles between different photo-$z$ bins are obvious. This reflects that the lensing has a broader kernel in the redshift dimension. We summarise the SNRs in Tab.\ref{snr} and the reduced $\chi^2$ in Tab. \ref{rechi2}. The total SNR$\simeq15$ and $22$ in the ``4 modules*yr'' and ``48 modules*yr'' cases, respectively. The reduced $\chi^2\simeq1$ suggests no significant deviation between the model and the data and validates that our spectrum estimation binning choices are appropriate.   
Finally, we summarize our simulation and spectrum estimation pipeline in the cartoon picture, Fig. \ref{ppl}.

\section{Cosmological constraints from cosmic shear-CMB lensing cross-correlation}\label{sec5}

The final step is to study the cosmological implications of the shear-CMB cross-correlation signal. We estimate the cosmological parameter constraint ability by using the Markov Chain Monte Carlo (MCMC) method. 
We use the \verb'Emcee' code \citep{Foreman_Mackey_2013}, a public implementation of the affine invariant MCMC ensemble sampler \citep{goodman2010ensemble}. We assume a Gaussian likelihood functions for the cross-spectrum 
\eqsali{
-2\log\mathcal{L}(\hat{D}^{XY}_\nu|\boldsymbol{\theta})=\chi^2
=\dsum_{\nu\nu'}\left(\hat{D}^{XY}_\nu-D^{XY}_\nu(\boldsymbol{\theta})\right)^T\Tilde{\mathbb{C}}_{\nu\nu'}^{-1}\left(\hat{D}^{XY}_{\nu'}-D^{XY}_{\nu'}(\boldsymbol{\theta})\right)\;,
}
where $\boldsymbol{\theta}$ represents
the set of parameters, including the cosmological as well as the nuisance parameters. $\mathbb{C}$ is the covariance matrix. $\hat{D}^{XY}_\nu(\boldsymbol{\theta})$ and $D^{XY}_\nu(\boldsymbol{\theta})$ are the measured bandpower and weighted theoretical spectra, $D^{XY}_\nu(\boldsymbol{\theta})$ is defined as
\eqsali{
D^{XY}_\nu(\boldsymbol{\theta})=\dsum_{i=1}^{\Delta L}w(\nu,\ell_i)D(\ell^{i}_\nu,\boldsymbol{\theta})
}
where $w(\nu,\ell_i)$ are the weights of bandpower $L$, we have chosen equal weights and normalized the weights for all bandpowers $\dsum_{i=1}^{\Delta L}w(\nu,\ell_i)=1$. The priors are summarized in Tab. \ref{prior}. The posterior on the model parameters is then given by
\eqsali{
\mathcal{P}(\boldsymbol{\theta}|\hat{D}^{XY}_\nu) =\mathcal{L}(\hat{D}^{XY}_\nu|\boldsymbol{\theta})\mathcal{P}(\boldsymbol{\theta})\;,
}
where $\mathcal{P}(\boldsymbol{\theta})$ are the priors.

\begin{table}
    \centering
\begin{tabular}{ccc}
\hline
Parameters & Fiducial value &Prior\\
\hline
$\Omega_m$      &$0.314$ &$(0.05,0.7)$\\
$h$             &$0.67$  &fixed\\
$\Omega_b$      &$0.049$ &fixed\\
$\sigma_8$      &$0.811$ &$(0.3,1.3)$\\
$n_s$           &$0.96$  &fixed\\
$A_{\rm IA}$    &$1.0$   &$(-5,5)$\\
\hline
$\Delta^1_z$    &$0.005$ &fixed\\
$\Delta^2_z$    &$0.005$ &fixed\\
$\Delta^3_z$    &$0.005$ &fixed\\
$\Delta^4_z$    &$0.005$ &fixed\\
$\sigma_z$      &$0.05$  &fixed\\
\hline
\end{tabular}
    \caption{The cosmological and nuisance parameters used in our simulation. The fiducial values are adopted from Planck-2018 \citep{aghanim2020planck} and COSMOS-2015 \citep{2018MNRAS.480.2178C}.}
    \label{prior}
\end{table}

In this work, we are interested in the $\Omega_m$ and $\sigma_8$ constraints. Moreover, we convert the $\sigma_8$ constraint into $S_8$. \footnote{We do not present here the constraints by adding the redshift bias since we find it has negligible effect.} For $S_8$, we use the following definition
\eqsali{
S_8=\sigma_8\left(\dfrac{\Omega_m}{0.3}\right)^\alpha\;,
}
where, the power law index $\alpha$ is calculated via the PCA method \citep{abdi2010principal}.
\begin{table*}
\centering
\begin{tabular}{ccccc}
\hline
\quad & I& II& III& IV\\
\hline
\vspace{5pt}
Data&N04+photo-$z$   & N04+photo-$z$+IA&  N048+photo-$z$ & N048+photo-$z$+IA\\
Model & shear & shear+IA\\
\hline
\end{tabular}
    \caption{Summary of the data vectors and model templates used in the cosmological constraint.}
\label{tab:datavector}
\end{table*}

\begin{figure}
    \centering
    \includegraphics[width=0.8\columnwidth]{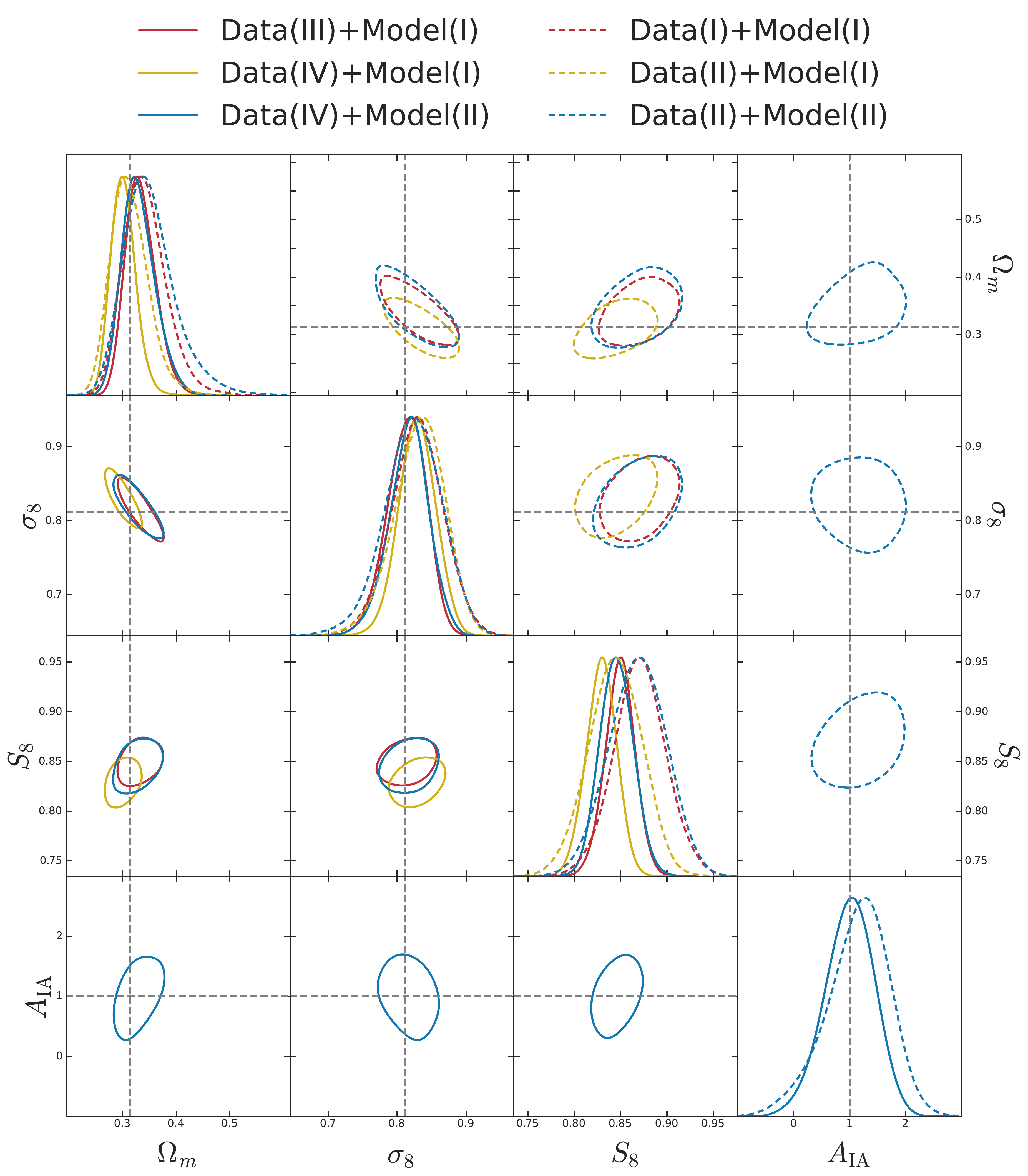}
    \caption{The constraint results of $\sigma_8$, $\Omega_m$, $S_8$ and $A_{\rm IA}$. $1\sigma (68.3\%)$ C.L. are shown. }
    \label{contour_N04}
\end{figure}

For the analysis, we considered four types of data and two types of theoretical models. The data comprises various kinds of noises and biases, while the model difference lies in whether the intrinsic alignment is included or not. We summarized our data vectors and model templates in Table \ref{tab:datavector}. MCMC chains were run by combining the aforementioned data and model templates. Our findings indicate that the typical 1$\sigma$ errors on $\sigma_8$ are approximately $0.038$ with the AliCPT-1 ``4 modules*yr'' setup and approximately $0.027$ with the AliCPT-1 ``48 modules*yr'' setup. Additionally, we discovered that if the IA is included in the data but not in the fitting template, it can introduce some level of bias in estimating $\sigma_8$ estimation. Our main results for parameter estimation are summarized in Table \ref{conlim} and Figure \ref{contour_N04}. Notably, the most significant shift in Figure \ref{contour_N04} is the contour from Data-IV with Model-I (solid yellow), emphasizing the importance of adequately modeling intrinsic alignment.

\begin{table*}
    \centering
\begin{tabular}{ccccccc}
\hline
 Parameter         &Data(I)+Model(I)                     & Data(II)+Model(I)              &Data(II)+Model(II)                &Data(III)+Model(I)                     & Data(IV)+Model(I)               &Data(IV)+Model(II) \\
\hline
\vspace{5pt}
{\boldmath$\Omega_m       $}   & $0.342^{+0.032}_{-0.044}   $& $0.313^{+0.029}_{-0.039}   $& $0.351^{+0.036}_{-0.055}   $& $0.334^{+0.024}_{-0.031}   $& $0.301^{+0.020}_{-0.025}   $& $0.330^{+0.026}_{-0.035}   $\\
{\boldmath$\sigma_8       $}   & $0.829\pm 0.038            $& $0.832^{+0.040}_{-0.033}   $& $0.823^{+0.043}_{-0.038}   $& $0.814^{+0.031}_{-0.026}   $& $0.829\pm 0.027            $& $0.817^{+0.030}_{-0.027}   $\\
{\boldmath$S_8            $}   & $0.871\pm 0.028            $& $0.844\pm 0.029            $& $0.868\pm 0.031            $& $0.849\pm 0.016            $& $0.829\pm 0.016            $& $0.846\pm 0.018            $\\
{\boldmath$A_{\rm IA}     $}   &  /                          &  /                          & $1.14^{+0.63}_{-0.47}      $&  /                          &  /                          & $0.999^{+0.47}_{-0.41}     $\\
{\boldmath$\alpha$}            & $0.40                      $& $0.41                      $& $0.37                      $& $0.42                      $& $0.42                      $& $0.39$                      \\
\hline
\end{tabular}
    \caption{The parameter estimation results of $\sigma_8$, $\Omega_m$, $S_8$, $A_{\rm IA}$. $1\sigma (68.3\%)$ C.L. are shown. The results of $S_8$ are derived from $\sigma_8$ and $\Omega_m$ posteriors. The power law index $\alpha$ is calculated via the PCA method \citep{abdi2010principal}.}
    \label{conlim}
\end{table*}

As for the power index in $S_8$, we compute it from the posterior directly. In detail, the power law index $\alpha$ was obtained from the correlation matrix of $\sigma_8$ and $\Omega_m$ in logarithmic space with Perform principal component analysis (PCA) algorithm (e.g. \citealt{abdi2010principal}), which gives the eigenvectors and eigenvalues for the normalized variables by using \verb'getdist' code \citep{2019arXiv191013970L}.

\begin{figure}
    \centering
    \includegraphics[width=\columnwidth]{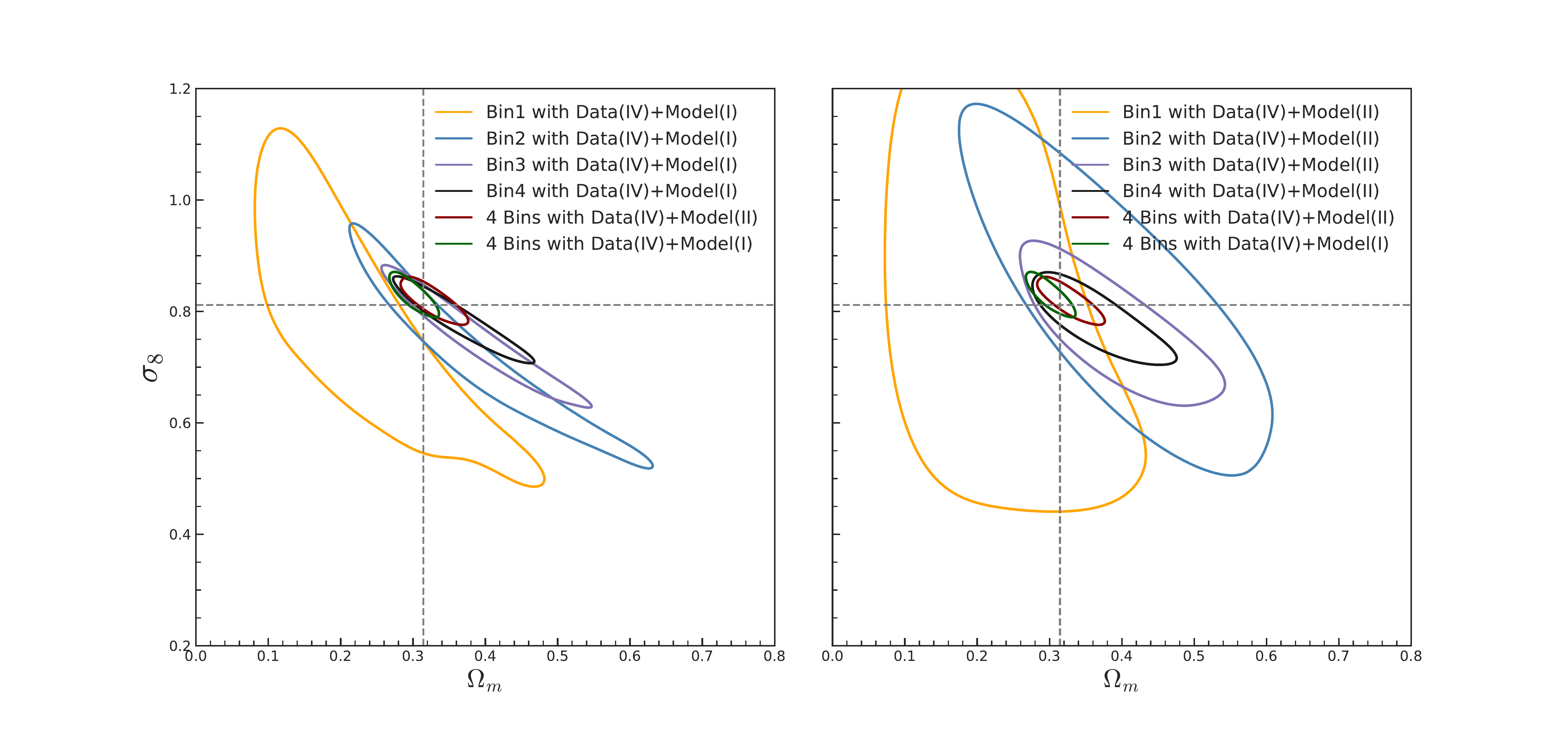}
    \caption{The constraint results of $\sigma_8$ vs. $\Omega_m$ with N048 noise in different photo-$z$ bins. $1\sigma (68.3\%)$ C.L. are shown. 
    The left panels show the results without properly considering the intrinsic alignment in the template fitting; while the right panels show the correct one. The input parameter $\sigma_8$ and $\Omega_m$ is marked by gray dashed lines.}
    \label{GI-G}
\end{figure}

The aim of this study is to investigate the intrinsic alignment bias on $\sigma_8$ and identify its reasons. We analyze the parameter constraints derived from individual photo-$z$ bins to achieve this goal, and the resulting findings are presented in Fig. \ref{GI-G}. The left panels of Fig. \ref{GI-G} illustrate the outcomes without considering intrinsic alignment in the template fitting process, while the right panels display the corrected results. The dark red and dark green contours in both the left and right panels represent the same values, shown for comparison purposes. It is observed that the first photo-$z$ bin deviates significantly from the combined constraint in the left panel, with a simultaneous decrease in $\sigma_8$ and $\Omega_m$.  However, after properly correcting for intrinsic alignment bias, the contours of each individual photo-$z$ bin overlap in the direction of constant-$S_8$,  which corresponds to the anti-diagonal direction in the $\sigma_8-\Omega_m$ plane). 

We conducted a comparative analysis of the constraining power with and without the inclusion of the correct IA model, which introduces additional nuisance parameters. To achieve this, we calculated a Figure of Merit (FoM). The FoM is defined as follows:
\eqsali{
{\rm FoM}=\dfrac{1}{\sqrt{{\rm det}[\mathbb{C}(\sigma_8,\Omega_m)]}}
}
where $\mathbb{C}$ refers to the (parameter) covariance matrix between
$\sigma_8$ and $\Omega_m$, the results are shown in Tab. \ref{tab:FoM}.

\begin{table*}
\centering
\begin{tabular}{ccccccc}
\hline
\quad& Data(I)+Model(I)&Data(II)+Model(I)&Data(II)+Model(II)&  Data(III)+Model(I)&Data(IV)+Model(I)&Data(IV)+Model(II)\\
\hline
\vspace{2pt}
FoM&991.45&1094.84&738.12&2404.07&2773.41&1942.53\\
\hline
\end{tabular}
    \caption{The figure-of-merit of each data vector and model template.}
\label{tab:FoM}
\end{table*}

\section{Conclusion}

In this work, we explore the cosmological constraints for the future CSST $\times$ AliCPT. We construct simulated maps for cosmic shear and CMB lensing 
based on the experimental nominal setup. In order to evade the complicated ray tracing technique, in this paper, we developed a simulation pipeline based on the Gaussian realization of the given signal and noise spectra. 
We forecast the S/N and the constraints on the cosmological parameters for the cross-correlation, considering statistical error from the two observations. We study the impact of the most important lensing systematics, photo-$z$ error, photo-$z$ bias, intrinsic alignment, and multiplicative bias, on the predicted cosmological parameters.

More specifically, we simulate the maps (Fig.\,\ref{k_g_maps}) according to CSST and AliCPT-1 nominal parameter setup. We consider standard shape noise for CSST cosmic shear (Fig.\,\ref{NeGI}), and the N0 noise from the disconnected primary CMB for AliCPT (Fig.\,\ref{Nphi}). As the map-building method (Eq.\,\eqref{eq build map} and \cite{kamionkowski1997statistics}) is based on Gaussian random fields, the covariance of their cross-correlation will contain the contribution for the above noises and the cosmic variance. We note that for AliCPT CMB lensing, the noise varies for the ``4 modules*yr'' and ``48 modules*yr'' stages. We perform the standard Pseudo-$C_\ell$ spectrum estimation and find the shear-CMB cross-correlation can reach the SNR$\simeq15$ for the ``4 modules*yr'' case, and the SNR$\simeq22$ for the ``48 modules*yr'' case, as shown in Fig.\,\ref{pseudo-Cls}. We investigate the cosmological implication of these cross-correlated signals in Fig.\,\ref{contour_N04}. We find that for the ``4 modules*yr'' case, the typical 1$\sigma$ errors on $\sigma_8$ is about $0.038-0.043$; for the ``48 modules*yr'' case, the typical 1$\sigma$ errors on $\sigma_8$ is about $0.027-0.030$, which is promising in investigating the current $S_8$ tension.

As an extension, we also explore the impact of photo-$z$ bias, multiplicative bias and intrinsic alignment, which are the main sources of systematics in weak lensing. In the generated mock data, we shift the mean redshift to represent the photo-$z$ bias and input an intrinsic alignment signal following the NLA model (Eq.\,\eqref{eq NLA IA} and Fig.\,\ref{NeGI}). 
We note that the true IA signal could potentially deviate from the assumed NLA model, for example, the TATT model \citep{Blazek2019,Samuroff2021,Hoffmann2022,Blazek2015,Samuroff2019,Samuroff2022} or the halo model \citep{Fortuna2021}. We leave those alternatives to future studies, as they are more dominant at smaller scales, while in this work the limit from CMB lensing noise (see Fig.\,\ref{Nphi}) reduce their impact. Using additional observables to self-calibrate the impact from IA is an alternative solution \citep{Yao2023}.
We show the contamination of photo-z and intrinsic alignment in the observed power spectra in Fig.\,\ref{pseudo-Cls}. We find that for the required photo-$z$ precision for CSST with $\Delta_z=0.005$, the bias in the power spectrum is negligible, while the IA contamination with $A_{\rm IA}=1$ is more significant. We find that if we do not consider the intrinsic alignment in the spectrum modeling, this will introduce about $0.6\sigma$ shift in $\sigma_8$ but an almost negligible effect on $S_8$ (Fig.\,\ref{contour_N04}). By including the correct IA model while introducing more nuisance parameters, the figure-of-merit in the $\sigma_8-\Omega_m$ space will be reduced from $\simeq2404$ to $\simeq1942$ (Fig.\,\ref{GI-G}), representing the loss in the cosmological constraining power to the IA parameter.
As for the multiplicative bias, we investigate its impact with the Fisher matrix method. With the typical value ($\sigma_m=0.01$) of the stage-III survey, the CMB lensing-cosmic shear cross-correlations are very insensitive to the multiplicative bias.

Interestingly, the map-making method of this paper provides not only an alternative check to the conventional Fisher matrix method but can also quickly generate correlated maps. 
A similar method has already been applied to the DES Year 3 joint analysis of galaxy clustering and weak lensing \cite{krause2021dark}.
This technique is essential in discussing systematic contaminations when combined with future simulations, as we can directly use maps from simulations rather than assume a model for the power spectrum, especially when sometimes the simulation and the model deviate at some level \citep{Jagvaral2022,Schneider2019}. However, there are some caveats that shall be properly addressed. First of all, the result presented here is based on the flat-sky and limber approximation, which obstructs us to use the full overlapped area of the two surveys. Hence, it limits us to reveal the full power of the cross-correlation. To do so, we need to use the curved sky expression and spherical harmonic transformation instead of the Fourier transformation. Second, we only vary three major parameters, namely $\sigma_8$, $\Omega_m$, and $A_{\rm IA}$ due to the limited SNR. Compared with the cosmic shear auto-correlations, the cross-correlation is still only playing a sub-leading role in the cosmology implication. But as we show in the validation section, it is immune to some systematic auto-correlation, such as the multiplicative bias.
Besides, we note that it is also important to include the impact of non-Gaussian covariance and other sources of systematics, such as the baryonic effect, but they are beyond the scope of this work and we leave them for future studies.


\section*{Data Availability}

The data underlying this article will be shared on reasonable request to the corresponding author.

\section*{Acknowledgements}

BH and ZYW are supported by the China Manned Space Project with No.CMS-CSST-2021-B01 and the National Natural Science Foundation of China Grants No. 11973016. JY acknowledges the support of the China Postdoctoral Science Foundation (2021T140451). XKL is supported by NSFC of China under Grant No. 11933002 and No. U1931210, and No. 12173033. ZHF and DZL acknowledge the support from NSFC under 11933002, U1931210, and from China Manned Space Project with No.CMS-CSST-2021-A01. DZL is also supported by the NSFC grant 12103043.
Some of the results in this paper have been derived using the \verb'healpy' \citep{Zonca:2019vzt}, \verb'HEALPix' \citep{2005ApJ...622..759G}, \verb'pyccl' \citep{Chisari_2019}, \verb'NaMaster' \citep{10.1093/mnras/stz093}, \verb'Emcee' \citep{Foreman_Mackey_2013} and \verb'getdist' \citep{2019arXiv191013970L} packages.

\clearpage
\bibliographystyle{mnras}
\bibliography{refs} 




\appendix
\section{Validation}
In order to check the algorithm and test the constraint of the parameters, we use the Fisher method to make a purely theoretical cosmological constraint, and we compute the theoretical covariance. The angular power spectra are computed within the multipoles range $(0<\ell<800)$ and binned $\Delta \ell=40$, the Gaussian covariance is calculated by

\begin{figure}
    \centering    
    \includegraphics[width=\columnwidth]{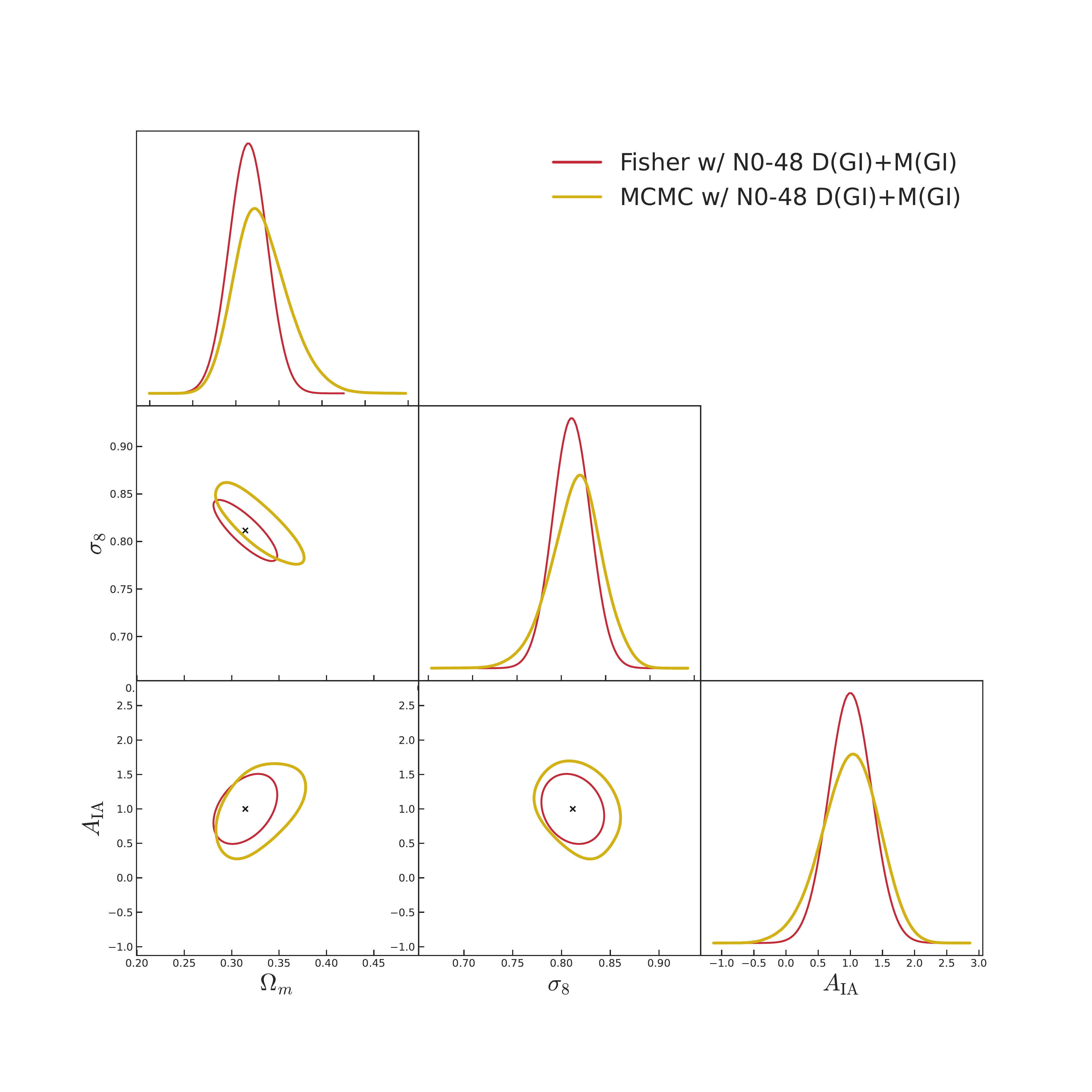}
    \caption{MCMC-Fisher matrix comparison. Red contours denote constraint results from Fisher's methods and the yellow are from the MCMC method, black crosslike marks denote the fiducial value of each parameter. The data input in Fisher's method is purely theoretical.}
    \label{Fisher-MC}
\end{figure}

\eqsali{
\mathbb{C}_{\nu\nu',{\rm theo}}=\dfrac{\delta_{LL'}}{(2L+1)\Delta Lf_{\rm sky}}\left[C^{XY}_{\nu}C^{XY}_{\nu'}+(C^{XX}+N^{XX})(C^{YY}_{\nu\nu'}+\delta_{\nu\nu'}N^{YY}_{\nu\nu})\right]\;,
}
where $\delta_{LL'}$ is the Kronecker delta function $C^{XY}_{\nu}\;,C^{XX}\;,C^{YY}_{\nu\nu'}$ are theoretical power spectra. Fisher matrix is written as follows
\eqsali{
\mathbb{F}_{pp'}=\left(\derivp{C^{XY}_{\nu}}{p}\right)\mathbb{C}^{-1}_{\nu\nu',{\rm theo}}\left(\derivp{C^{XY}_{\nu'}}{p'}\right)^T\;,
}
and the covariance matrix of parameters
\eqsali{
\mathbb{C}_{pp'}=\mathbb{F}^{-1}_{pp'}\;,
}
where $p \in \{\Omega_m,\sigma_8.A_{\rm IA}\}$. For comparison, the theoretical error bars were obtained from the Fisher matrix. They are in generally good agreement with the MCMC error estimates, which are shown in Fig.\,\ref{Fisher-MC}, however, are slightly larger (by up to $\sim 20\%$).

\begin{figure}
    \centering
    \includegraphics[width=\textwidth]{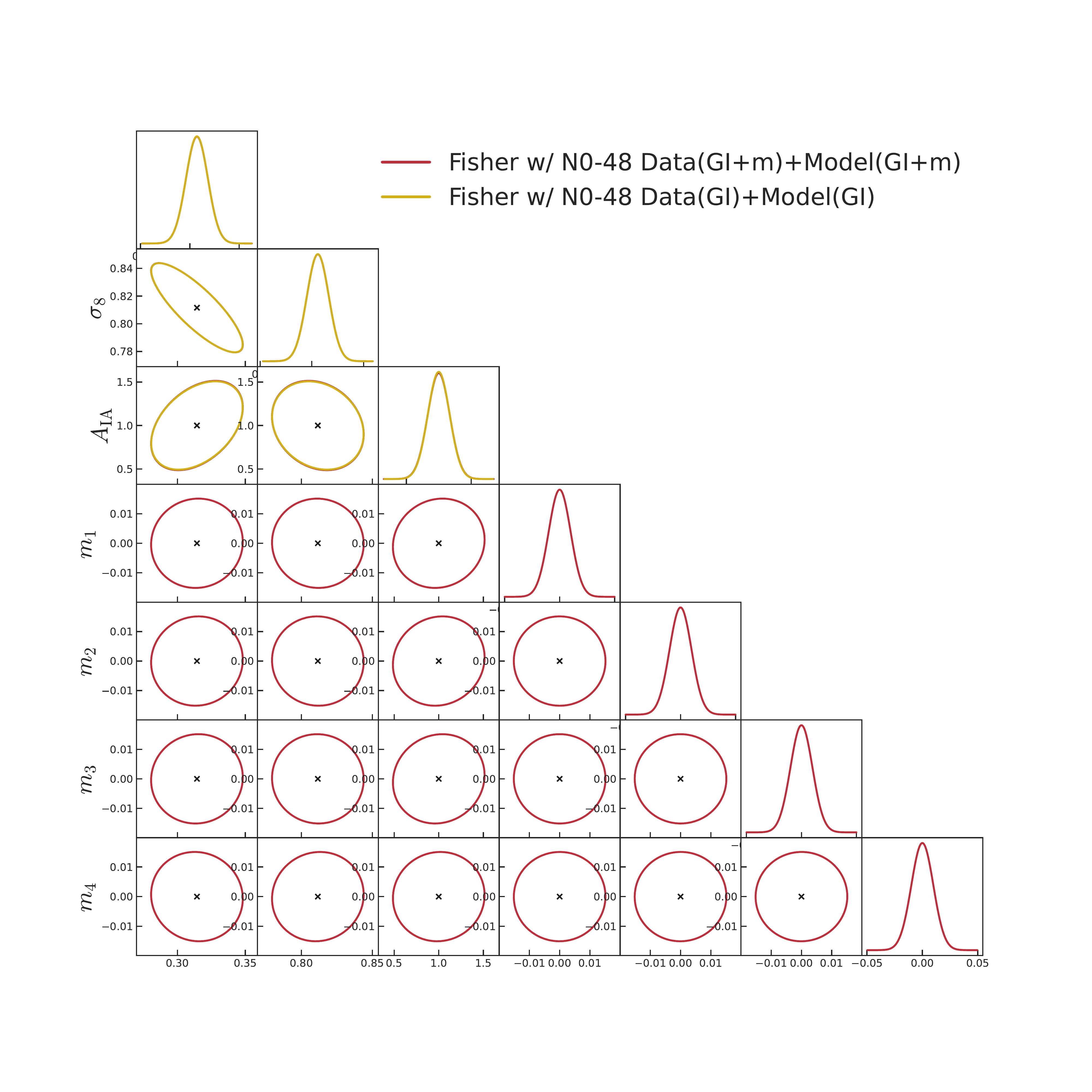}
    \caption{Validation with multiplicative bias. Red contours denote constraint results from Fisher's methods without multiplicative bias, and the yellow contours are from Fisher's methods with multiplicative bias}
    \label{FisherIA_m}
\end{figure}

\begin{figure}
    \centering
    \includegraphics[width=\textwidth]{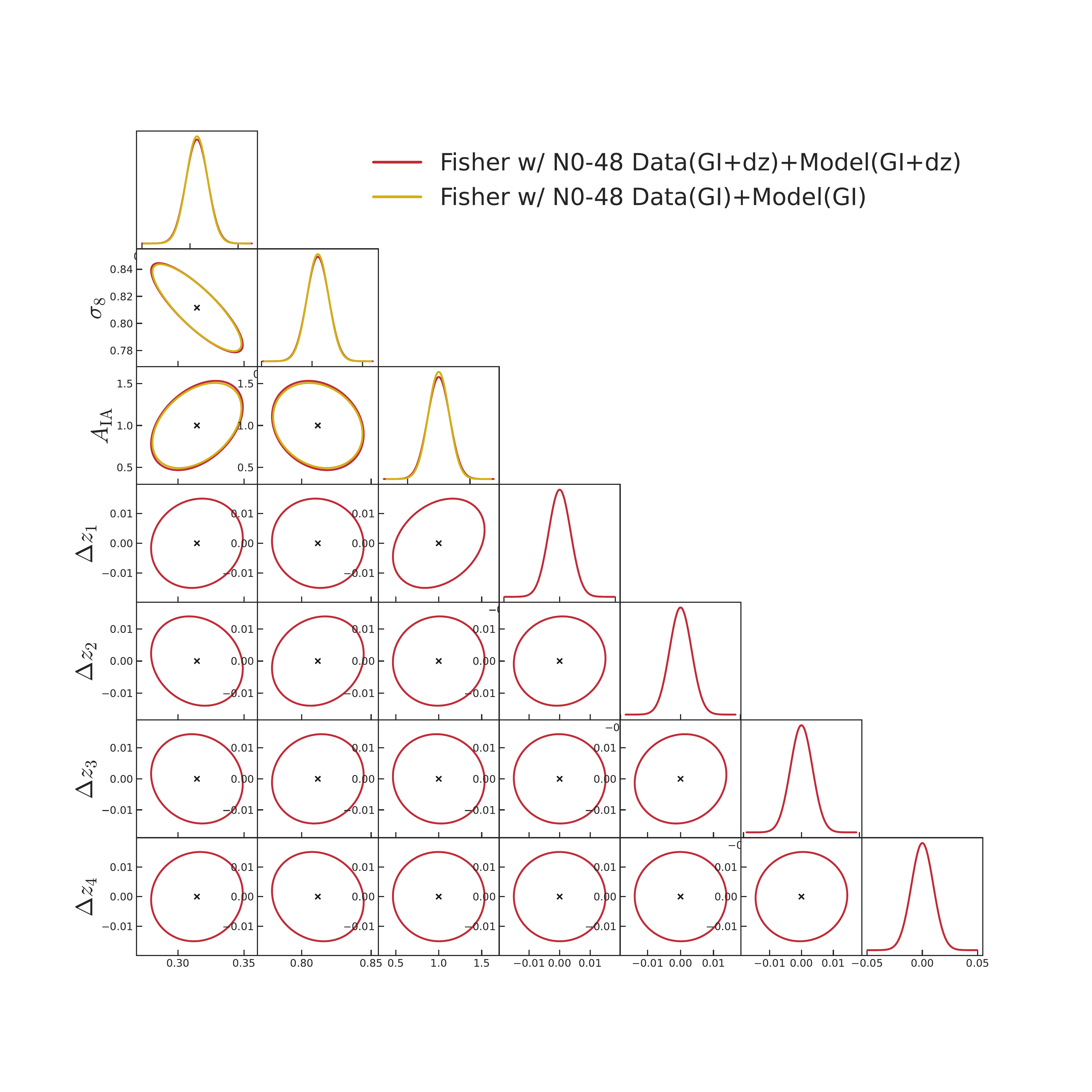}
    \caption{Validation with photo-$z$ bias. Red contours denote constraint results from Fisher's methods without photo-$z$ bias, and the yellow contours are from Fisher's methods with photo-$z$ bias}
    \label{FisherIA_dz}
\end{figure}

We consider the bias from galaxy shape measurement in different photo-$z$ bins and it can be conventionally described approximately as a linear model with an additive bias $c$ and a multiplicative bias
$m$ \citep{10.1093/mnras/sts454,10.1093/mnras/stv781}.
\eqsali{
\gamma^{\rm obs}=(1+m)\gamma^{\rm true}+c\;,
}
here we only consider the effect of multiplicative bias $m$, parameter $p \in \{\Omega_m,\sigma_8.A_{\rm IA},m_1,m_2,m_3,m_4\}$ and adopt $c=0$, and the corresponding theoretical cross spectrum should be written as
\eqsali{
\Tilde{C}^{XY}_\nu=(1+m_\nu)C^{XY}_\nu\;,
}
the subscript $\nu$ of $m$ denotes multiplicative bias in different photo-$z$ bins. We adopt a Gaussian prior with null mean and standard deviation $\sigma_m=0.01$ of multiplicative bias in our validation, the results are shown in Fig.\,\ref{FisherIA_m}.
And we also consider the effect of photo-$z$ bias in different photo-$z$ bins, parameter $p \in \{\Omega_m,\sigma_8.A_{\rm IA},\Delta_{z1},\Delta_{z2},\Delta_{z3},\Delta_{z4}\}$, we adopt a Gaussian prior with null mean and standard deviation $\sigma_{\Delta z}=0.01$, and the results are shown in  Fig.\,\ref{FisherIA_dz}.
We note that both multiplicative bias and photo-$z$ bias have a small impact on the constraints of cosmological parameters.


\bsp	
\label{lastpage}
\end{document}